\documentclass[aps,prb,twocolumn,floatfix,footinbib,superscriptaddress]{revtex4}

\usepackage{graphicx}
\usepackage{dcolumn}
\usepackage{bm}
\usepackage{xcolor}

\usepackage{ulem}
\usepackage{amsmath,amssymb}

\begin{document}

\title{Suppression of the optical linewidth and spin decoherence of a quantum spin center in a  \textit{p}--\textit{n} diode}

\author{Denis R. Candido}
 \email{denisricardocandido@gmail.com}
\affiliation{Department of Physics and Astronomy, University of Iowa, Iowa City, Iowa 52242,
USA}
\affiliation{Pritzker School of Molecular Engineering, University of Chicago, Chicago, Illinois 60637, USA}

\author{Michael E. Flatt\'e}%
 \email{michael\_flatte@mailaps.org}
\affiliation{Department of Physics and Astronomy, University of Iowa, Iowa City, Iowa 52242,
USA}
\affiliation{Pritzker School of Molecular Engineering, University of Chicago, Chicago, Illinois 60637, USA}
\affiliation{Department of Applied Physics, Eindhoven University of Technology, P.O. Box 513, 5600 MB, Eindhoven, The Netherlands}

\date{\today}

\begin{abstract} 
We present a quantitative theory of the suppression of the optical linewidth due to charge fluctuation noise in a \textit{p}--\textit{n} diode, recently observed in Anderson \textit{et al.}, Science \textbf{366}, 1225 (2019). We connect the local electric field with the voltage across the diode, allowing  for the identification of the defect depth from the experimental threshold voltage.
Furthermore, we show that an accurate description of the decoherence  of such spin centers 
requires a complete spin--1 formalism that yields a bi-exponential decoherence  process, and predict how reduced charge fluctuation noise suppresses the spin center's decoherence rate.






\end{abstract}

\maketitle


\section{Introduction}

The role of the environment on the optical linewidth and spin decoherence of an optically-accessible spin center is well-known to be significant, and a major step forward in reducing environmental effects has been achieved recently by placing a spin center in a semiconductor \textit{p}--\textit{n} diode wherein the effects of charge fluctuations can be suppressed\cite{anderson2019electrical}. 
Considerable effort has already been devoted to optimize such optically-accessible quantum-coherent spin centers associated with defects in semiconductor crystals, including by reducing the sources of magnetic~\cite{electric-magnetic1,electric-magnetic2,electric-magnetic3,electric-magnetic4,magneticnoise1,magneticnoise2,magneticnoise3,magneticnoise4,magneticnoise5,magneticnoise6,magneticnoise7} or electrical~\cite{electric-magnetic1, electricnoise1,electric-magnetic2,electric-magnetic3,electric-magnetic4,electricnoise1,electricnoise2,electricnoise3} noise from nearby surfaces or other defects. A \textit{p}--\textit{n} diode, however, provides the possibility of dynamically removing sources of noise as well as studying the properties of noise under exceptionally controlled conditions. The interest in suppressing the optical linewidth and spin decoherence of these quantum-coherent  centers stems in part from their broad applicability  to  quantum information sciences, including quantum sensing~\cite{PhysRevX.10.011003,electric-magnetic1,schirhagl2014nitrogen,dolde2014nanoscale,RevModPhys.89.035002,RevModPhys.89.035002} and quantum memory~\cite{fuchs2011quantum}. Their discrete optical transitions, and their direct or indirect coupling to spin, provide an avenue to probe fundamental quantum properties, \textit{e.g.}, teleportation~\cite{hensen2015loophole} and spin-photon entanglement~\cite{robledo2011high,awschalom2018quantum}. These spin centers can also provide components for quantum networking, \textit{e.g.} through optical~\cite{togan2010quantum,de2010universal,Wehnereaam9288,awschalom2018quantum} or magnonic means~\cite{lukaprx,andrich2017long,PhysRevA.97.052303,PhysRevB.99.195413,PhysRevB.99.140403,lee2020nanoscale,candido2020predicted,neuman2020nanomagnonic}.  Divacancies in SiC~\cite{lowther1977vacancies,falko2014,seo2016,miao2019electrically,de2017stark,ivady2015theoretical}  and negatively-charged nitrogen-vacancy  (NV)    centers~\cite{doherty2013nitrogen,casola2018probing,awschalom2018quantum}  in  diamond both provide excellent optical addressability and long-lived spin coherence~\cite{seo2016,miao2019electrically,anderson2019electrical}, however the construction of electrical devices from SiC is often much simpler and less expensive than diamond devices. 
\begin{figure}[t!]
\begin{center}
\includegraphics[clip=true,width=1\columnwidth]{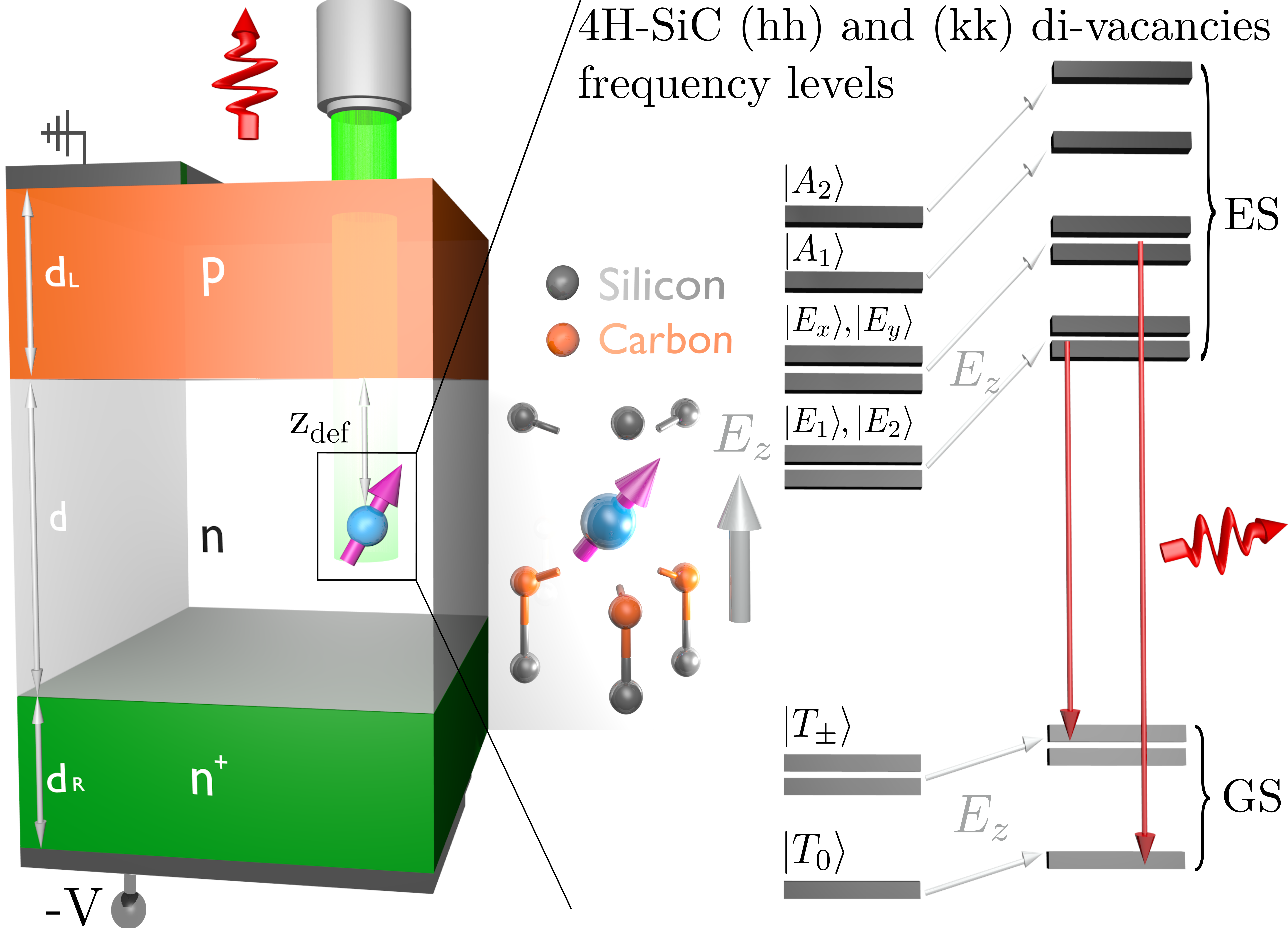}
\caption{Schematic view of the 4H-SiC \textit{p}--\textit{n}--\textit{n}$^+$ diode of Ref.~\onlinecite{anderson2019electrical} with one divacancy located in the \textit{n} region at $z=z_{\textrm{def}}$. The (hh) or (kk) divacancy lattice configuration is shown together with the response of its frequency to an applied electric field ($E_z$ along the defect symmetry axis) produced by the  diode voltage. The red arrows represent the PL transitions addressed in this work. } 
\label{fig1}
\end{center}
\end{figure}

 Here we present a quantitative theory describing the fundamental properties of the quantum-coherent spin center in the environment of a \textit{p}--\textit{n}--\textit{n}$^+$ diode (\textit{e.g.} Ref.~\onlinecite{anderson2019electrical}, and Fig.~\ref{fig1}), including the response of the optical transition energy, optical linewidth, and spin coherence to the diode voltage. In a complex and nonlinear fashion the macroscopic voltage across the device affects the local electric field (and thus the optical transition energy) as well as the fluctuating electric field both in the nearby region and in the electrical device contacts. The shift of the spin center's optical transition energies via the Stark effect~\cite{tamarat2006stark,electric-magnetic1,bassett2011electrical,acosta2012dynamic,klimov2014electrically,christle2017,de2017stark,miao2019electrically,anderson2019electrical} is expected in this configuration, however we show a rigorous,  quantitative and analytic connection between the experimentally applied voltage and the defect's transition energy, and from that we can determine the depth of the defect within the device ($z_{\textrm{def}}$ in Fig.~\ref{fig1}). In Ref.~\onlinecite{anderson2019electrical} both the optical transition energies and linewidths were observed via photoluminescence (PL). For new spin centers with known positions in such a diode our results can be used to determine the local electric field, which with measured optical transition energy shifts would determine the  electric dipole moments of  the ground and excited states of the new spin center. Furthermore, we propose improved diode designs to yield shifts in the emission frequencies in response to electric fields in excess of 1~THz for 4H-SiC divacancies, while avoiding  dielectric breakdown of the material. 
 
 The optical linewidth has been observed to narrow when a 
 diode is biased to deplete nearby electrical carriers. Here our quantitative theory for the charge fluctuation noise within both the depleted regions and the non-depleted regions (the densely doped contacts) produces an analytical formula for the optical linewidth as a function of the voltage, the defect position and the diode design (thicknesses of regions and doping densities). The analytic calculations are confirmed by Monte Carlo simulations of the charge fluctuation noise, and reproduce accurately the experimental results of Ref.~\onlinecite{anderson2019electrical}. We therefore propose altered diode designs, and spin center placement within diodes, which may further reduce the optical linewidth of similar quantum-coherent spin centers. 

Finally, we derive an extensive  theory for the ground state spin decoherence of defects with $C_{3v}$ point group symmetry, e.g., NV centers, (hh) and (kk) SiC divacancies. A key result that the correct coherence times can only be accurately calculated through a spin--1 formalism including the entire defect ground state subspace. We find a bi-exponential decoherence process that cannot be obtained from the spin-1/2 formalism. By considering the effect of charge fluctuation noise on the spin decoherence we predict the consequences for diode voltage for the quantum spin center's coherence times.  We obtain  analytical expressions for these coherence times as a function of the diode voltage, diode densities, temperature and the defect's position within the diode. These results identify a potential enhancement of the coherence time when the defect is within the depletion region, although this conclusion depends on the magnitude of the characteristic charge fluctuation time and on the ground state dipole moment. For the diode and defect parameters  of Ref.~\cite{anderson2019electrical} no spin coherence time enhancement is obtained,  agreeing with the experimental data.

The presentation of these results in this article is begins, in Sec.~\ref{defect-sec} with the description of the ground state (GS) and excited state (ES) Hamiltonians for  defects with $C_{3v}$ and $C_{1h}$ point group symmetry (electronic structure shown schematically in Fig.~\ref{fig1}), including the coupling to an external electric field. In Sec.~\ref{diode-sec} we present a complete theory for the spatially-dependent electric field and charge density in \textit{p}--\textit{n} and \textit{p}--\textit{n}--\textit{n$^+$} diodes as a function of  voltage, focusing on the experimentally-relevant regime of reverse voltage (charge depletion).  The spin center's optical emission spectrum, including the optical linewidth calculated  analytically and via numerical Monte Carlo simulation of the charge fluctuation noise is presented in Sec.~\ref{defect-emission}.  Finally, in Sec.~\ref{deco-sec} we predict the variation of the coherence times of defects with $C_{3v}$ symmetry as a function of diode voltage and design.

\section{Spin Center Hamiltonians}
\label{defect-sec}

The electronic structure of defects with point group symmetries $C_{3v}$ and $C_{1h}$ are strongly modified by the presence of an electric field. Examples of defects with $C_{3v}$ point group symmetry are the negatively-charged nitrogen-vacancy (NV) in diamond~\cite{loubser1978electron,van1990electric,tamarat2006stark,hossain2008ab,de2010universal,togan2010quantum,bassett2011electrical,maze2011,electric-magnetic1,doherty2011negatively,acosta2012dynamic,doherty2012,doherty2013nitrogen,dolde2014,schirhagl2014nitrogen,dolde2014nanoscale,rogers2015singlet,ivady2015theoretical,seo2016} and both (hh) with (kk) divacancies in 4H-SiC~\cite{lowther1977vacancies,klimov2014electrically,ivady2015theoretical,seo2016,de2017stark,christle2017,miao2019electrically,anderson2019electrical}. Defects with $C_{1h}$ symmetry include we have (hk) and (kh) divacancies in 4H-SiC~\cite{lowther1977vacancies,klimov2014electrically,ivady2015theoretical,seo2016,de2017stark,christle2017,miao2019electrically,anderson2019electrical}. The Stark effect~\cite{tamarat2006stark,electric-magnetic1,bassett2011electrical,acosta2012dynamic,klimov2014electrically,christle2017,de2017stark,miao2019electrically,anderson2019electrical} is responsible for coupling these discrete defect energy levels to the electric field. For  this situation, the total ground state (GS) Hamiltonian in the triplet basis $\left|T_{-}\right\rangle ,\left|T_{0}\right\rangle ,\left|T_{+}\right\rangle$ for defects with $C_{3v}$ point group symmetry is~\cite{loubser1978electron,van1990electric,tamarat2006stark,hossain2008ab,de2010universal,togan2010quantum,bassett2011electrical,maze2011,electric-magnetic1,doherty2011negatively,acosta2012dynamic,doherty2012,doherty2013nitrogen,dolde2014,schirhagl2014nitrogen,dolde2014nanoscale,rogers2015singlet,ivady2015theoretical,seo2016,lowther1977vacancies,klimov2014electrically,ivady2015theoretical,seo2016,de2017stark,christle2017,miao2019electrically,anderson2019electrical}

\begin{align} 
\frac{{\cal H}_{\rm{GS}}^{3v}}{h} & =D_{\rm{G}}\left(S_{z}^{2}-\frac{2}{3}\right)+d_{\rm{G}}^{\parallel}E_{z}\left(S_{z}^{2}-\frac{2}{3}\right) \nonumber \\
 & +d_{\rm{G}}^{\perp} E_{x}\left(S_{y}^{2}-S_{x}^{2}\right)+d_{\rm{G}}^{\perp} E_{y}\left(S_{x}S_{y}+S_{y}S_{x}\right)\label{hgs},
\end{align}
where $h$ is  Planck's constant, $D_{\rm{G}}$ is the zero energy splitting between the triplet states $m_s=0$ and $m_s=\pm 1$, $\textbf{S}$ are the triplet spin-1 matrices, $\textbf{E}=(E_x,E_y,E_z)$ is the electric field and $d_{\rm{G}}^{\parallel},d_{\rm{G}}^{\perp}$ are electric dipole constants, and for $C_{3v}$ point group symmetry $d_{\rm{G}}^{\parallel}\neq d_{\rm{G}}^{\perp}$. Here the $z$ direction corresponds to the defect symmetry axis. The eigenfrequencies are \begin{align}
    E_0 &= -\frac{2}{3}(D_{\rm{G}}+d_{\rm{G}}^{\parallel}E_z),\\
    E_{\pm} &= \frac{1}{3}\left(D_{\rm{G}}+d_{\rm{G}}^{\parallel}E_z\right) \pm |d_{\rm{G}}^{\perp}|\sqrt{E_x^2+E_y^2},
\end{align}
where we see that $E_z$ enters within the diagonal matrix elements, whereas $E_{x,y}$ couples the $\left|T_{-}\right\rangle ,\left|T_{+}\right\rangle$ subspace, lifting its initial degeneracy. For the excited state (ES) the Hamiltonian is ~\cite{maze2011,doherty2011negatively,doherty2013nitrogen,christle2017}
\begin{align}
\frac{{\cal H}_{\rm{ES}}^{3v}}{h} & =D_{\rm{E}}^{\parallel} \left(S_{z}^{2}-\frac{2}{3}\right)-\lambda_{\rm{E}}^{\parallel}\sigma_{y}\otimes S_{z}\nonumber \\
 & +D_{\rm{E}}^{\perp}\left[\sigma_{z}\otimes\left(S_{y}^{2}-S_{x}^{2}\right)-\sigma_{x}\otimes\left(S_{y}S_{x}+S_{x}S_{y}\right)\right]\nonumber \\
 & +\lambda_{\rm{E}}^{\perp}\left[\sigma_{z}\otimes\left(S_{x}S_{z}+S_{z}S_{x}\right)-\sigma_{x}\otimes\left(S_{y}S_{z}+S_{z}S_{y}\right)\right]\nonumber \\
 & +d_{\rm{E}}^{\parallel}E_{z}+d_{\rm{E}}^{\perp}\left(\sigma_{z}E_{x}-\sigma_{x}E_{y}\right)\label{hes},
\end{align}
where $\sigma_{x,y,z}$ are the Pauli matrices, $\lambda$'s ($D_E$'s) are the parameters due to the spin-orbit (spin-spin) interaction, and $d_E$'s represents the electric dipole moments. Although there is no analytical form for the ES frequencies of Eq.~(\ref{hes}), in this work we are only interested in their response to $E_z$. As the term $d_{\rm{E}}^{\parallel}E_z$ enters only in the diagonal of Eq.~(\ref{hes}), it yields a constant shift for the whole ES frequency subspace. In Fig.~\ref{fig1} we draw schematically the GS and ES discrete frequency levels and their response to an applied $E_z$ electric field, with red arrows representing the spin conserving optical transitions addressed in this work. 

For defects with $C_{1h}$ symmetry, e.g., (kh) or (hk) divacancies in 4H-SiC, the Hamiltonians for the ground and excited states have the same form
~\cite{miao2019electrically}
\begin{align}   
\frac{{\cal H}_{\rm{GS}(\rm{ES})}^{1h}}{h}  &=D_{\rm{G}(\rm{E})}^z\left(S_{z}^{2}-\frac{2}{3}\right)+\tilde{d}_{\rm{G}(\rm{E})}^{\parallel}E_{z}\left(S_{z}^{2}-\frac{2}{3}\right) \nonumber \\
& +D_{\rm{G}(\rm{E})}^x\left(S_{y}^{2}-S_{x}^{2}\right)+D_{\rm{G}(\rm{E})}^y\left(S_{x}S_{y}+S_{y}S_{x}\right)\label{hgs-1h},
\end{align}
with $D_{\rm{G}(\rm{E})}^x = \tilde{d}_{\rm{G}(\rm{E})}^{\perp} E_{x} +D_{\rm{G}(\rm{E})}^0$ and $D_{\rm{G}(\rm{E})}^y=\tilde{d}_{\rm{G}(\rm{E})}^{\perp} E_{y} +D_{\rm{G}(\rm{E})}^0$. In contrast to the defect GS Hamiltonian with $C_{3v}$ symmetry [Eq.~(\ref{hgs})], here we have a lifting of the degeneracy between $m_s=\pm 1$ triplet states in the absence of an electric field. This is due to the appearance of the $D_{\rm{G}(\rm{E})}^0$ crystal fields terms due to the reduced defect symmetry, yielding
\begin{align}
    E_{\tilde{A}_0\left(A_{0}'\right)} &= -\frac{2}{3}(D_{\rm{G}(\rm{E})}^z+\tilde{d}_{\rm{G}(\rm{E})}^{\parallel}E_z),\\
    E_{{\tilde{A}_{\pm}\left(A_{\pm}'\right)}} &= \frac{1}{3}\left(D_{\rm{G}(\rm{E})}^z+\tilde{d}_{\rm{G}(\rm{E})}^{\parallel}E_z\right) \pm \sqrt{\left(D_{\rm{G}(\rm{E})}^x\right)^2+\left(D_{\rm{G}(\rm{E})}^y\right)^2}.
\end{align}

\section{$p$--$n$ and $p$--$n$--$n^+$ diode electric fields and carrier densities}
\label{diode-sec}

In this work our defects are assumed to be placed or built within a \textit{p}--\textit{n}--\textit{n}$^+$ diode configuration. Within the reverse bias regime, where a negligible current passes through the diode, the dc electric field experienced by our defect is produced by an interplay of the electric field arising from the depletion region formation and from the voltage $-V$ applied across the diode device [Fig.~\ref{fig1}]. The orange and green regions in Fig.~\ref{fig1} represent the \textit{p} and \textit{n}$^+$ regions, respectively, with $N_A$ acceptor and $N_D$ donor impurity densities. The white region represents the \textit{n} region, with donor density $N\ll N_D$. In the following subsections we use the theory of diodes for the reverse bias regime to derive analytically the important  quantities for the defect's optical and spin dynamical properties, e.g., carrier densities and electric field profiles. Secondly, we plot and analyze these quantities for the  diode configuration of Ref.~\cite{anderson2019electrical} under different applied voltages and at different positions.

\subsection{Fundamental theory and key equations}

Diodes usually consist of a homogeneous and neutral semiconductor with a spatially dependent doped region [Fig.~\ref{fig2}(a)]. The semiconductor's band gap, $E_g=\epsilon_c-\epsilon_v$, where $\epsilon_c$ and $\epsilon_v$ correspond to the energy of the conduction and valence bands respectively [Fig.~\ref{fig2}(a)]. We consider first the $p$--$n$ region of the $p$--$n$--$n^+$ diode shown in Fig.~\ref{fig1}, and so we will use $N$ for the donor concentration in the $n$ region, and $N_D$ for the donor concentration in the $n^+$ region. By doping the material with an acceptor impurity density $N_A$ for $z<0$ (with energy $\epsilon_a \gtrsim \epsilon_v$), and a donor impurity density $N$ for $0<z<d$ (with energy $\epsilon_d \lesssim \epsilon_c$), we obtain the  \textit{p}--\textit{n} diode region (orange and white regions within Fig.~\ref{fig1}). For temperatures $T$ such that $\epsilon_c-\epsilon_d \gtrsim  k_B T$ and $\epsilon_a-\epsilon_v \gtrsim  k_B T$, where $k_B$ is the Boltzmann constant, the dopants  are excited and populate the conduction and valence bands with electrons and holes [Fig.~\ref{fig2}(a)]. After ionization the carriers in the conduction and valence bands are free to move and start to recombine with each other. This recombination produces a region with few free carriers (depletion region) and a spatially dependent charged background that in turn produces an electric field along the $z$ direction. 

For the fully ionized case, we have an approximate background charge distribution $\rho(z)$ given by
\begin{equation}
\rho\left(z\right)=\begin{cases}
0 & z<-\tilde{d}_{p}(V)\\
-eN_{A} & -\tilde{d}_{p}(V)<z<0\\
eN & 0<z<\tilde{d}_{n}(V)<d \label{rho} \\
0 & z>d
\end{cases},
\end{equation}
where $e>0$ is the fundamental electronic charge, and the positions $-\tilde{d}_p(V)$ and $\tilde{d}_n(V)$ define the spatial boundaries of the depletion region. The background charge density produces both an electrostatic potential $\phi(z)$ and an electric field $\textbf{E}=-({\partial\phi}/{\partial z}) \hat{z}$, obtained through the Poisson equation
\begin{equation}
    \partial_{z}^{2}\phi\left(z\right)=-\frac{1}{\epsilon }\rho\left(z\right) \label{poisson},
\end{equation}
where $\epsilon$ is the dielectric constant of our diode material.
Here we assume small variations of the electrostatic potential along $x$ and $y$ axis, so that $\partial_{x}^2\phi=\partial_{y}^2\phi \approx 0$, and Eq.~(\ref{poisson}) follows. Using the boundary condition  $\left.\phi\right|_{z\rightarrow-\infty}=\left.{d\phi}/{dz}\right|_{z\rightarrow-\infty}=\left.{d\phi}/{dz}\right|_{z\rightarrow\infty}=0$, we solve Eq.~(\ref{poisson}), obtaining 
\begin{equation}
\phi\left(z\right)=\begin{cases}
0 & z<-\tilde{d}_{p}(V)\\
\frac{e}{2\epsilon}N_{A}\left[z+\tilde{d}_{p}(V)\right]^{2} & -\tilde{d}_{p}(V)<z<0\\
\phi_{\infty}\left(V\right)-\frac{e}{2\epsilon}N\left[z-\tilde{d}_{n}(V)\right]^{2} & 0<z<\tilde{d}_{n}(V)\\
\phi_{\infty}\left(V\right) & \tilde{d}_{n}(V)<z
\end{cases},\label{phi-pn}
\end{equation}
with
\begin{align}
\tilde{d}_{n}\left(V\right) & =\sqrt{\frac{2\text{\ensuremath{\epsilon}}\phi_{\infty}\left(V\right)}{e}\frac{N_{A}/N}{\left(N_{A}+N\right)}} \leq d,\label{dn1} \\ 
\tilde{d}_{p}\left(V\right) & =\frac{N}{N_{A}}\tilde{d}_{n}\left(V\right). \label{dp1}
\end{align}
Here $\phi_{\infty}\left(V\right)=\frac{1}{e}\left\{E_{g}+k_{B}T\ln\left[\frac{N_{D}N_{A}}{N_{c}\left(T\right)P_{v}\left(T\right)}\right]-eV\right\}$ is obtained by fixing a constant chemical potential $\mu$  along the entire sample, with $N_{c}\left(T\right)=\frac{1}{4}\left(\frac{2m_{c}k_{B}T}{\pi\hbar^{2}}\right)^{3/2}$ and $P_{v}\left(T\right)=\frac{1}{4}\left(\frac{2m_{v}k_{B}T}{\pi\hbar^{2}}\right)^{3/2}$, and effective conduction and valence band masses, $m_{c}$, and $m_{v}$, respectively. The electric field is straightforwardly obtained from Eq.~(\ref{phi-pn}),
\begin{equation}
E_z\left(z,V\right)=\begin{cases}
0 & z<-\tilde{d}_{p}(V)\\
-\frac{e}{\epsilon}N_{A}\left[z+\tilde{d}_{p}(V)\right] & -\tilde{d}_{p}(V)<z<0\\
\frac{e}{\epsilon}N\left[z-\tilde{d}_{n}(V)\right] & 0<z<\tilde{d}_{n}(V)\\
0 & \tilde{d}_{n}(V)<z<d \label{electric-pn}
\end{cases}.
\end{equation}
To obtain Eq.~(\ref{electric-pn}) we  assume the voltage drops only inside the depletion region, so the electric field vanishes outside. Finally, the majority carrier densities of electrons and holes can be expressed as a function of the position $z$
and the voltage $V$, as 
\begin{align}
n_{c}\left(z,V\right) & =\begin{cases}
0 & z<0\\
{2N}/\left[e^{-\frac{\phi_{\infty}\left(V\right)-\phi\left(z\right)}{k_{\rm{B}}T}} +1\right] & 0<z<d\\
{2N_{D}}/\left[e^{-\frac{\phi_{\infty}\left(V\right)-\phi\left(z\right)}{k_{\rm{B}}T}}+1\right] & z>d \label{nc}
\end{cases},
\end{align}
and
\begin{align}
p_{v}\left(z,V\right)=\begin{cases}
{2N_{A}}/\left(e^{\frac{\phi\left(z\right)}{k_{\rm{B}}T}}+1\right) & z<0\\ 
0 & 0<z<d\\
0 & z>d \label{pv}
\end{cases}.
\end{align}
Here we have neglected the minor carrier contributions $\approx \frac{n_{c}^2}{N_D}$ and $\approx \frac{p_{v}^2}{N_A}$, as they are much smaller as compared to the majority carriers Eqs.~(\ref{nc}) and (\ref{pv}). Hence, we do not expect them to have a major influence in our results.


We note that for the critical voltage $V_c$, defined through $\tilde{d}_n(V_c)=d$ [Eq.~(\ref{dn1})], with solution
\begin{equation}
    V_c=-\frac{e d^2}{2\epsilon}\frac{N\left(N_{A}+N\right)}{N_{A}}+E_{g}+k_{B}T \thinspace \textrm{ln}\left[\frac{N_{A}N_{D}}{N_{c}(T)P_{v}(T)}\right], \label{Vc}
\end{equation}
we achieve full depletion of the \textit{n} region. Therefore, for $V<V_c$ the corresponding Eqs. (\ref{rho}) and (\ref{phi-pn})--(\ref{electric-pn}) for the effective \textit{p}--\textit{n} diode do not hold, and must be replaced by the corresponding equations for the \textit{p}--\textit{n}--\textit{n}$^+$ diode.  For this situation, the background charge density is
\begin{align}
    \rho\left(z\right)=\begin{cases}
0 & z<-d_{p}(V)\\
-eN_{A} & -d_{p}(V)<z<0\\
eN & 0<z<d\\
eN_{D} & d<z<d+d_{n}(V)\\
0 & z>d+d_{n}(V)
\end{cases},
\end{align}
where the positions $-{d}_p$ and ${d}_n$ define the new spatial boundaries of the depletion region and read
\begin{align}
d_{p}\left(V\right) & =d\frac{N-N_{D}}{N_{A}+N_{D}} \label{dp-pin} \\
 +& \sqrt{\frac{N_{D}}{N_{A}}\frac{2\text{\ensuremath{\epsilon}}\phi_{\infty}\left(V\right)}{2\left(N_{A}+N_{D}\right)}+d^{2}\frac{N_{D}}{N_{A}}\frac{\left(N_{D}-N\right)\left(N_{A}+N\right)}{\left(N_{A}+N_{D}\right)^{2}}}, \nonumber
\end{align}
\begin{align}
d_{n}\left(V\right) & =-d\frac{N+N_{A}}{N_{A}+N_{D}} \label{dn-pin}\\
 +&\sqrt{\frac{N_{A}}{N_{D}}\frac{2\text{\ensuremath{\epsilon}}\phi_{\infty}\left(V\right)}{e\left(N_{A}+N_{D}\right)}+d^{2}\frac{N_{A}}{N_{D}}\frac{\left(N_{D}-N\right)\left(N_{A}+N\right)}{\left(N_{A}+N_{D}\right)^{2}}}. \nonumber
\end{align}
The electrostatic potential is obtained through Poisson's equation Eq.~(\ref{poisson}), yielding
\begin{equation}
    \phi\left(z\right)=\begin{cases}
0 & z<-d_{p}(V)\\
\frac{e}{2\epsilon}N_{A}\left[z+d_{p}(V)\right]^{2} & -d_{p}(V)<z<0\\
\frac{eN_A}{\epsilon}\left[\frac{d_{p}^{2}\left(V\right)}{2}+d_{p}\left(V\right)x\right]-\frac{eN}{2\epsilon}z^{2} & 0<z<d\\
\phi_{\infty}\left(V\right)-\frac{e}{2\epsilon}N_{D}\left[z-d_{n}(V)\right]^{2} & d<z<d+d_{n}(V)\\
\phi_{\infty}\left(V\right) & z>d+d_{n}(V) \label{phi}.
\end{cases}
\end{equation}
The electric field within our diode is straightforward determined 
\begin{equation}
{E}_z\left(z,V\right)=\begin{cases}
0 & z<-d_{p}(V)\\
-\frac{e}{\epsilon}N_{A}\left[z+d_{p}(V)\right] & -d_{p}(V)<z<0\\
-\frac{e}{\epsilon}N_{A}d_{p}(V)+\frac{e}{\epsilon}Nz & 0<z<d\\ \label{electric-pin}
\frac{e}{\epsilon}N_{D}\left[z-d_{n}(V)\right] & d<z<d+d_{n}(V)\\
0 & z>d+d_{n}(V)
\end{cases}
\end{equation}

\subsection{Results for various diode configurations}
\label{pin-results}
As motivated earlier, the idea of this work is to use the diode setup to manipulate and control the defect PL linewidth and frequencies and the coherence time of our defect. A good setup is achieved by setting $N\ll N_A \lesssim N_D$ with $d\gg d_L$, which leads to small charge concentration within a large spatial region, and a large electric field within the $N$ region. To simulate a realistic system, we use in this work the following diode parameters  from Ref.~\onlinecite{anderson2019electrical} that considers divacancies within a 4H-SiC $p$--$n$--$n^+$ diode, with $N_A=7\times 10^{18}$~cm$^{-3}$, $N \approx 4\times 10^{15}$~cm$^{-3}$, $N_D=10\times 10^{18}$~cm$^{-3}$, $d_R \gg d=10$~$\mu$m, $d_L=400$~nm, $\epsilon=9.6\epsilon_0$ and $T\approx10$~K. Although the results here are presented for the particular case of 4H-SiC diode, the same would hold for any diode material with alterations in the material parameters if the donors and acceptors have shallow states. Some materials, such as diamond, do not possess both shallow donors and acceptors, and so the expressions here would be considerably more complicated to account for incomplete dopant ionization.

\begin{figure}[t!]
\begin{center}
\includegraphics[clip=true,width=1\columnwidth]{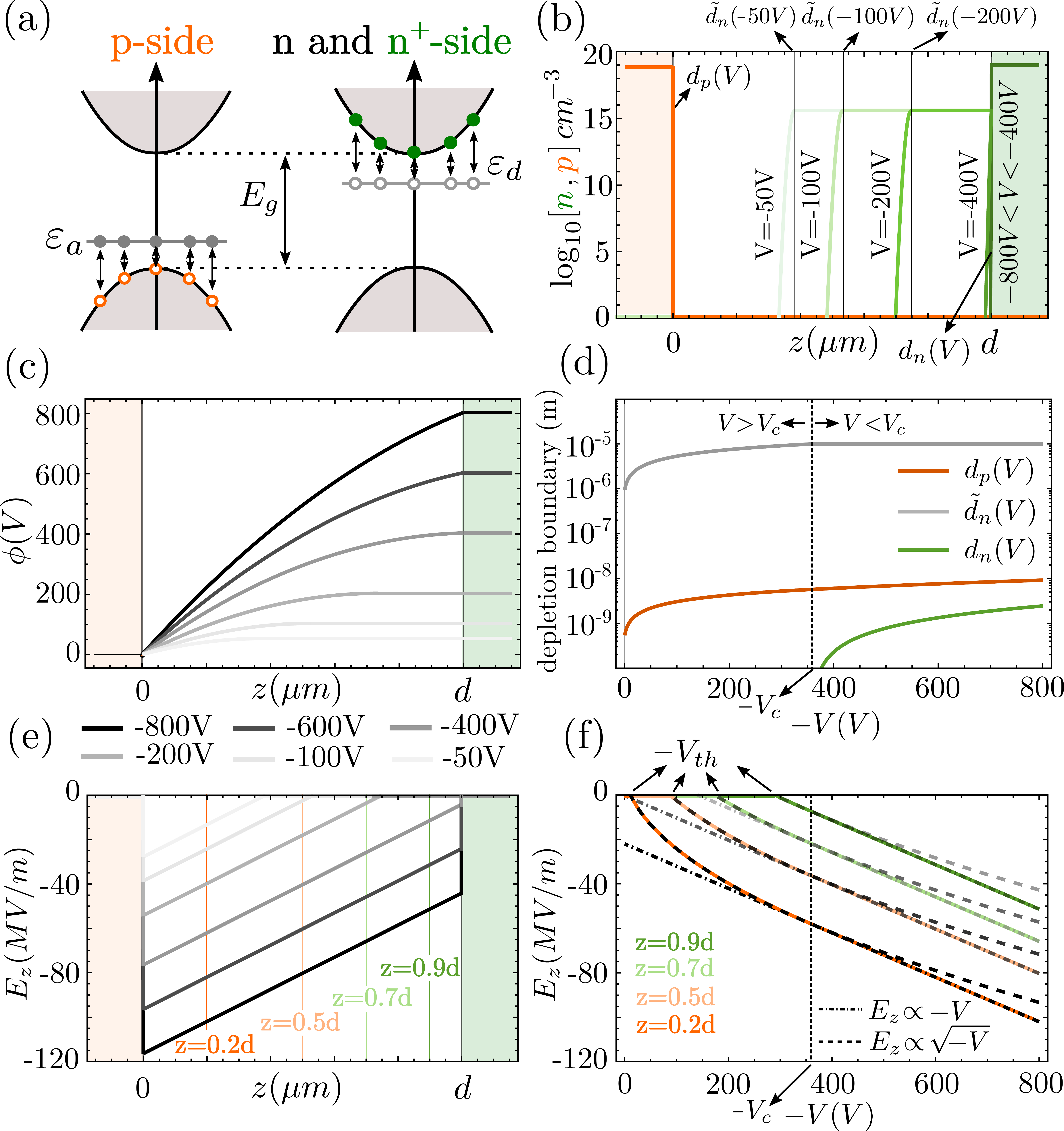}
\caption{(b) Electron (n) and hole (p) density carriers along $z$ direction for different applied reverse voltages $V$. (c) Electrostatic potential $\phi(z)$ along the diode for voltages spanning from $V=-12$~V to $V=-400$~V. (d) Depletion region boundary positions as a function of $V$. (e) Electric field profile along $z$ direction for the voltages spanning from $V=-12$~V to $V=-400$~V. (f) Electric field as a function of $V$ at four different positions $z=0.2d$, $z=0.5d$, $z=0.7d$ and $z=0.9d$.} 
\label{fig2}
\end{center}
\end{figure}

All the results for the corresponding diode quantities are grouped within Fig.~\ref{fig2}. In Fig.~\ref{fig2}(b) we plot the density of free carriers Eqs.~(\ref{nc})--(\ref{pv}) for different voltages within the reverse bias regime, which shows that the larger the modulus of $V$ the more we deplete the charges carriers. The spatial boundary positions of the depletion region is plotted in Fig.~\ref{fig2}(d), which also captures the increase of the depleted region size as a function of the voltage. Moreover, for $|V|>|V_c|$, the \textit{n} region becomes 100$\%$ depleted and $\tilde{d}_n(V)$ becomes a constant with value $d$. In both Figs.~\ref{fig2}(d) and (f) we indicate the critical voltage $V_c \approx -\frac{e d^2}{2\epsilon}\frac{N\left(N_{A}+N\right)}{N_{A}}\approx -373$~V, in which the system stops behaving as an effective {\textit{p}--\textit{n}} diode, and start behaving as a {\textit{p}--\textit{n}--\textit{n}$^+$} diode. Fig.~\ref{fig2}(c) shows the electrostatic potential profile along z for different voltages spanning from $-50$ to $-800$~V. 
In Fig.~\ref{fig2}(e) we plot the electric field profile within the diode for different voltages in the reverse bias regimes. The electric field shows a linear trend with respect to the position until we reach the outside of the depletion region, in which the electric field becomes zero. The linear trend in the slope is easily understood through the integration of Gauss's equation [Eq.~(\ref{electric-pn})] $\partial_z E_z(z) = - {\rho(z)}/\epsilon$ $\rightarrow$ $E_z(z) \propto e N z /\epsilon$, while the amplitude of the electric field is proportional to the voltage drop $V/d$. 
In Fig.~\ref{fig2}(f) we plot the local electric field at positions $z=0.2d$, $0.5d$, $0.7d$ and $0.9d$ as a function of the applied voltages $V$. For large absolute values of $V$, a linear trend with respect to $V$ is observed for most of the voltage region, and is understood through $E_z\approx V/d$ [dashed-dotted line Fig.~\ref{fig2}(f)]. However, we also observe a non-linear trend appearing for $|V|<|V_c|$. To understand that we have to recall that for these voltages the \textit{n} region is not fully depleted, and its depletion region length depends on $V$ approximately as $\sqrt{\frac{-2\text{\ensuremath{\epsilon}}V}{eN}}$~[Eq.~(\ref{dn1})], thus yielding a non-linear dependence of the electric field with respect to $V$ given by $  -\frac{e}{\epsilon}N \left( z - \sqrt{\frac{-2\text{\ensuremath{\epsilon}}V}{eN}} \right)$ [dashed line Fig.~\ref{fig2}(f)]. Therefore, we understand that this non-linear trend on $E_z$ vs. $V$ is a peculiarity of an {\textit{p}--\textit{n}--\textit{n}$^+$} diode becoming an effective {\textit{p}--\textit{n}} diode. It is important to mention that the depletion regions sizes in both \textit{p} and {\textit{n}$^+$} regions, $d_{p}(V)$ and $d_n(V)$, also have a $\sqrt{-V}$ dependence as can be seen from Eqs.~(\ref{dp-pin}) and (\ref{dn-pin}), and in Fig.~\ref{fig2}(d). Therefore, in principle even for $|V|>|V_c|$ we would expect an electric field deviating from the linear trend. However, due to the large electronic density of both \textit{p} and {\textit{n}$^+$} regions $N_D,N_A\gg N$, we find $d_{p}(V),d_n(V) \ll d$, and hence we can assume the voltage drop $V$ happening only along $\approx d$, thus yielding $E_z \approx V/d$. Finally, the electric field at the fixed positions $z=0.2d$, $0.5d$, $0.7d$ and $0.9d$ in Fig.~\ref{fig2}(f) approach to zero at the threshold voltages $V_{th}$, for which these positions match the depletion boundary, i.e., $\tilde{d}_n(V_{th})=z$, thus experiencing no electric field. Interestingly, through this fact we can determine the defect's position along the $z$ axis, ${z}_{\rm{def}}$, by accessing only the experimental $V_{th}$ value coming from PL measurements. More specifically, when the defect is inside the depletion region, the shift of the PL frequency as a function of the voltage can be seen experimentally. However, when we tune $V=V_{th}$, the PL frequency stop responding to the applied voltage as the defect is now outside of the depletion region. This condition happens for
\begin{equation}
{z}_{\rm{def}}\left(V_{th}\right)	=\sqrt{\frac{2\text{\ensuremath{\epsilon}}}{e}\frac{N_{A}/N}{\left(N_{A}+N\right)}\phi_{\infty}\left(V_{th}\right)}, \label{zdef}
\end{equation}
and it allows for the precise determination of the spin center's position wtihin the diode. Finally, the different electric field trends as a function of the voltage are  important, since they establish the relation between the experimental applied voltage and the electric field felt by a defect located at $0<z_{\rm{def}}<d$. We summarize these trends using Eqs.~(\ref{dn1})--(\ref{electric-pn}) and (\ref{electric-pin}), with $N_A,N_D \gg N$ and $eV \gg E_g,k_BT$,
\begin{equation}
E_z\left(z_{\rm{def}},V\right)\approx\begin{cases}
0 & |V|<|V_{th}| \\
\frac{e}{\epsilon} N \left( z_{\rm{def}}-\sqrt{-\frac{2\epsilon V N_A/N}{e (N_A+N)}} \right) & |V_{th}|<V<|V_c|\\
\frac{e}{\ensuremath{\epsilon}}N\left(z_{\rm{def}}-\frac{d}{2}\right)+\frac{V}{d} & |V|>V_{c} 
\end{cases}\label{elecf}.
\end{equation}


\section{Defect spectrum emission}
\label{defect-emission}

In this section we establish and explore the relation between the defect optical emission spectrum (which can be measured, \textit{e.g.}, through photoluminescence) and the applied voltage across the diode. More specifically, we show analytically how the frequency and the linewidth of the defect PL depend on the reverse bias voltages. We also provide different schemes for the diode configurations and defect's position that yields THz shifts in the PL emission. Moreover, we compare our predictions with experimental data from Ref.~\onlinecite{anderson2019electrical} and good agreement is seen.

First, we assume the spin center is located at $\textbf{r}_{\rm{def}}=(x_{\rm{def}},y_{\rm{def}},z_{\rm{def}})$. From Hamiltonians Eqs.~(\ref{hgs}), (\ref{hes}) and (\ref{hgs-1h}), we then obtain the defect  transition frequencies as a function of the electric field. For the purpose of this work, we report the results corresponding to the PL of (hh), (kk) and (kh) 4H-SiC divacancies, which  were experimentally addressed in Ref.~\cite{anderson2019electrical}. They correspond to the $ \left|E_{y}\right\rangle  \rightarrow \left|T_{0}\right\rangle$ and $ \left|E_{1,2}\right\rangle \rightarrow \left|T_{\pm}\right\rangle $ transition for the (hh) and (kk) divacancies [Fig.~\ref{fig1}], and $\left|A_{0}^{'}\right\rangle \rightarrow \left|\tilde{A}_{0} \right\rangle$ transition for the (kh) divacancy, reading
 \begin{align}
{\Delta f^{3v}_{E_{y}\rightarrow T_{0}}}  & =\left(d_{E}^{\parallel}+\frac{2}{3}d_{G}^{\parallel}\right){E_{z}\left(z_{\rm{def}},V\right)}{}, \label{To-e}\\
{\Delta f^{3v}_{ E_{1,2}\rightarrow T_{\pm}}}   & =\left(d_{E}^{\parallel}-\frac{1}{3}d_{G}^{\parallel}\right){E_{z}\left(z_{\rm{def}},V\right)}{}, \label{Tpm-e12}\\
{\Delta f^{1h}_{A_{0}' \rightarrow \tilde{A}_0 }} & = -\frac{2}{3}\left(\tilde{d}_{E}^{\parallel}-\tilde{d}_{G}^{\parallel}\right)\cos(109.5^o) {E_{z}\left(z_{\rm{def}},V\right)}{}, \label{ao-ao}
\end{align}
where $E_{z}\left(z_{\rm{def}},V\right)$ is the electric field at $z_{\rm{def}}$ for voltage $V$ [Eq.~(\ref{elecf})] and $\cos(109.5^o)$ accounts for the decomposition of the electric field along the high symmetry axis of the (kh) divacancy. Although many works have studied the Stark effect in defects~\cite{tamarat2006stark,electric-magnetic1,bassett2011electrical,acosta2012dynamic,klimov2014electrically,christle2017,de2017stark,miao2019electrically,anderson2019electrical}, most of these were unipolar materials without significant charge depletion. None of them provided quantitative relation between the  voltage applied to a $p$--$n$ diode and the electric field experienced by the defect, which in turn is the microscopic quantity coupled to their energy levels. Here we obtain this relation [Eq.~(\ref{elecf})] by solving Poisson's equation [Eq.~(\ref{poisson})] for both $p$--$n$ and $p$--$n$--$n^+$ diodes. Most importantly, through Eqs.~(\ref{elecf})--(\ref{ao-ao}) we are able to understand the quantitative dependence of defect frequency shift on the voltage and the diode parameters. For instance, we are able to predict that as the $n$ region's doping density $N$ increases, the electric field at the defect also increases, which produces shifts to higher frequency  PL emission. Through this connection, it is possible to engineer better diodes in order to achieve higher frequency shifts using smaller voltages, which becomes important as the possible applied voltages reach limits determined by the dielectric breakdown field of the material. 

Due to the different dependence on the dipoles $d_{G}^{\parallel}$ and $d_{E}^{\parallel}$ of Eqs.~(\ref{To-e}) and (\ref{Tpm-e12}), it also becomes possible to determine both of the spin center dipole values from experimental measurements. Usually experimental measurements~\cite{tamarat2006stark,electric-magnetic1,bassett2011electrical,acosta2012dynamic,klimov2014electrically,christle2017,de2017stark,miao2019electrically,anderson2019electrical} only report the values of the effective dipole moments corresponding to the addressed PL transitions. Here we provide equations that, in principle, would allow the extraction of both the  $d_{G}^{\parallel}$ and $d_{E}^{\parallel}$  dipole moments. Through the experimental voltage dependence of the $\left|E_{x,y}\right\rangle \rightarrow \left|T_{0}\right\rangle$ and $ \left|E_{1,2}\right\rangle \rightarrow \left|T_{\pm}\right\rangle$ transitions, we obtain from Eqs.~(\ref{elecf}), (\ref{To-e}) and (\ref{Tpm-e12}), the experimental values for $d_{G}^{\parallel}$ and $d_{E}^{\parallel}$. Moreover, in what concerns the experimental work Ref.~\cite{anderson2019electrical}, we observe $\Delta f_{E_{y}\rightarrow T_{0}} \approx \Delta f_{E_{1,2}\rightarrow T_{\pm}}$, which results in $d_{E}^{\parallel}\gg d_{G}^{\parallel}$, thus showing that the dipole of the excited state manifold is the major property responsible for the Stark shift.

\begin{figure}[t!]
\begin{center}
\includegraphics[clip=true,width=1\columnwidth]{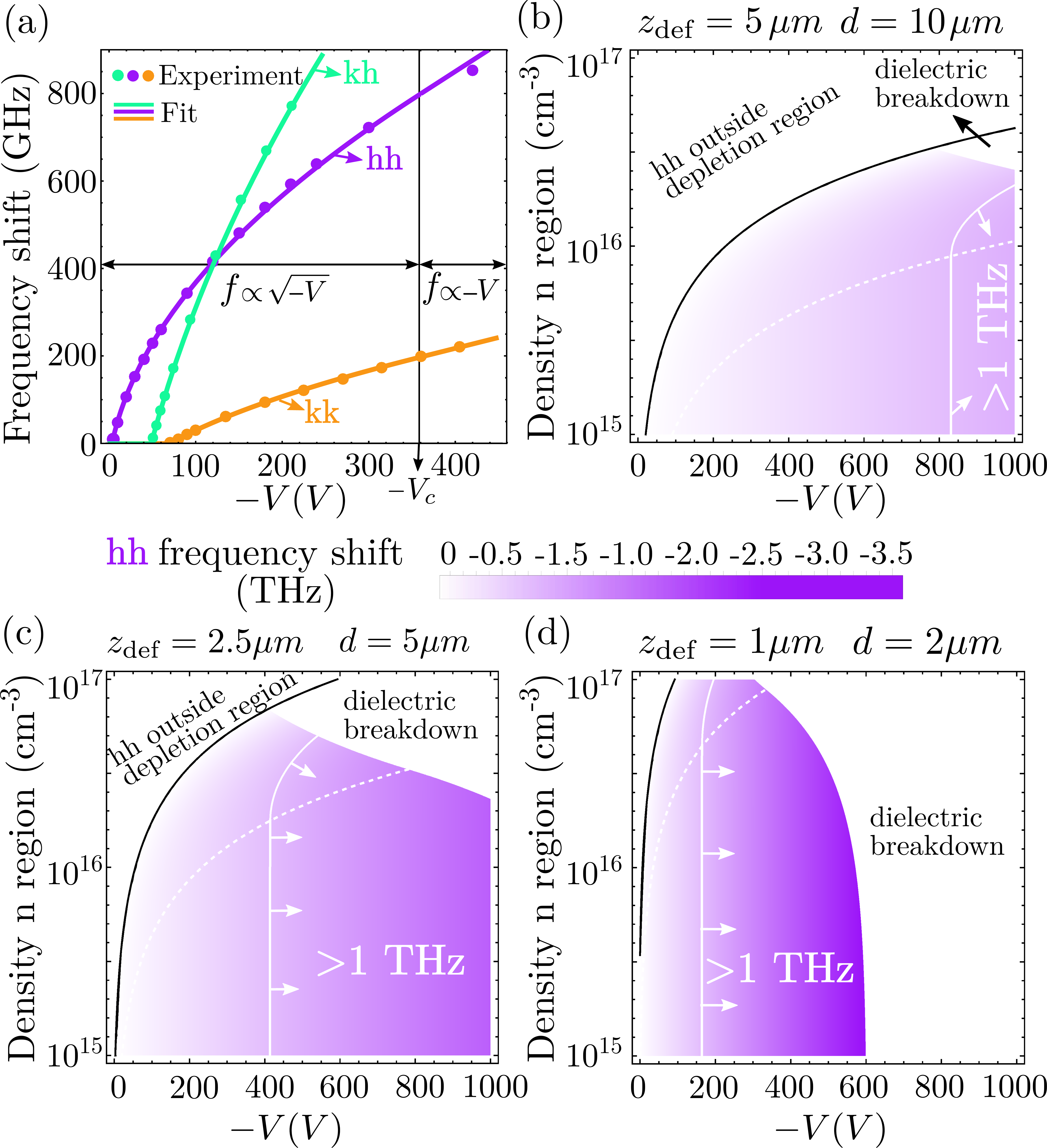}
\caption{(a) Fit for the frequency shift vs. voltage $V$ for the  $\left|E_{y}\right\rangle  \rightarrow \left|T_{0}\right\rangle$ and $ \left|E_{1,2}\right\rangle \rightarrow \left|T_{\pm}\right\rangle $ (hh) and (kk) transitions, and for the  $\left|A_{0}^{'}\right\rangle \rightarrow \left|\tilde{A}_{0} \right\rangle$ (kh) transition, assuming $N=4\times 10^{15}$~cm$^{-3}$. Frequency shift of the (hh) defect as a function of the voltage, and density of \textit{n} region for (b) $z_{\rm{def}}=d/2=5~\mu$m, (c) $z_{\rm{def}}=d/2=2.5~\mu$m and (d) $z_{\rm{def}}=d/2=1~\mu$m. The black solid line delimits the parameter region in which the defect is placed within and without the depletion region. The while solid line represtents $1$~THz frequency shift, and the dashed white line represents the change from $p$--$n$ to $p$--$n$--$n^+$ diode behavior.} 
\label{fig3}
\end{center}
\end{figure}

In Fig.~\ref{fig3}(a) we use Eqs.~(\ref{elecf})--(\ref{ao-ao}) to fit the experimental data of Ref.~\cite{anderson2019electrical} that contains the frequency shift of the (hh) and (kh) $ \left|E_{y}\right\rangle  \rightarrow \left|T_{0}\right\rangle$ and $ \left|E_{1,2}\right\rangle \rightarrow \left|T_{\pm}\right\rangle $ transitions, and of the (kh) $\left|A_{0}^{'}\right\rangle \rightarrow \left|\tilde{A}_{0} \right\rangle$  transition. We fit it assuming $d_G^{\parallel}=0$, as already justified before. The frequency shifts in Fig.~\ref{fig3}(a) have two different trends with respect to the voltages. For small voltage modulation, $|V|<|V_c|$, the frequenc{ies} ha{ve} a non-linear dependence on $V$, which was already explained in Sec.~\ref{pin-results} as resulting from the incomplete depletion of the \textit{n} region. Specifically, if the \textit{n} region is not completely depleted, the depletion length depends on the voltage as $\sqrt{-V}$ [Eq.~(\ref{dn1})], thus yielding a frequency shift $\propto E_z \approx -V/\sqrt{-V} \approx \sqrt{-V}$. For voltage modulation larger than $|V|>|V_c|$ the \textit{n} region becomes fully depleted and a frequency shift $\propto E_z \approx -V/d$ is expected, assuming the small depletion region extent into the \textit{p} and $n^+$ regions discussed previously. For the donor and acceptor density values reported in Ref.~\onlinecite{anderson2019electrical}, we obtain $V_c\approx-90$~V, which is not supported by the long non-linear trend within $0\lesssim |V| \lesssim 350$~V range of the experimental data [Fig.~\ref{fig3}(a)]. This suggest a larger unintentional doping density value than the reported $N\approx 1\times 10^{15}$~cm$^{-3}$ value~\cite{anderson2019electrical}. The best fit to the experimental data is presented in Fig.~\ref{fig3}(a), where we assume $N=4\times 10^{15}$~cm$^{-3}$. For this density, we obtain dipoles $d_{E,kk}^{\parallel}=4.95$~GHz/(MV/m), $d_{E,hh}^{\parallel}=12.75$~GHz/(MV/m) and $\tilde{d}_{E,kh}^{\parallel}-\tilde{d}_{G,kh}^{\parallel}=118.07$~GHz/(MV/m). While the (hh) and (kk) dipole values are in accordance with Refs.~\cite{anderson2019electrical,de2017stark,miao2019electrically}, the (kh) dipole value seems to deviate. This is understood as in the previous Refs.~\cite{anderson2019electrical,miao2019electrically}, the $109.5^o$ angle between the (hk) symmety axis and the direction of the electric field was not considered. In addition, the discrepancy may also be related to the highly anisotropic Stark shift. Furthermore, using Eq.~(\ref{zdef}) we are also able to determine the positions of  spin centers along the z axis by accessing the threshold voltages $V_{th}$, yielding $z_{\rm{def}}^{kk}=4.42$~$\mu m$, $z_{\rm{def}}^{hh}=1.32$~$\mu m$ and $z_{\rm{def}}^{kh}=3.53$~$\mu m$. For completeness, in Appendix~\ref{appendixc} we also provide the data fit using densities  $N=1\times 10^{15}$,  $2\times 10^{15}$ and $3\times 10^{15}$~cm$^{-3}$, which clearly shows worse agreement.

In Fig.~\ref{fig3}(b), (c) and (d) we plot the frequency shift for the (hh) divacancy $ \left|E_{y}\right\rangle  \rightarrow \left|T_{0}\right\rangle$ transition as a function of the voltage $V$ and the density of the \textit{n} region, $N$. We chose three different combinations of \textit{n} region length and defect's position: $d=10$~$\mu$m with $z_{\rm{def}}=5~\mu$m [Fig.~\ref{fig3}(b)], $d=5$~$\mu$m with $z_{\rm{def}}=2.5~\mu$m [Fig.~\ref{fig3}(c)] and  $d=2$~$\mu$m with $z_{\rm{def}}=1~\mu$m [Fig.~\ref{fig3}(d)]. In all of these three configurations we are able to obtain (hh) frequency shift $>1~$THz under operable voltages. However, as $E_z\approx V/d$, the smaller the length $d$, the smaller the applied voltage to observe Terahertz shifts. For the situation of Fig.~\ref{fig3}(d), we obtain Terahertz shifts even with small applied voltages $V\approx -200$~V. As we are going to discuss below, the only drawback of having diodes with small length $d$ is that defect becomes closer to the non-depleted \textit{p} and $n^+$ regions, thus being more sensitive to the electric noise caused by the fluctuation of the electric charges. In Figs.~\ref{fig3}(b), (c) and (d), the black solid lines separate the parameter space regions in which the defect is inside and outside the depletion region. The white solid lines delimit the parameter region where we have frequency shifts $>1$~THz, and the dashed white lines represent the parameter space region where our effective $p$--$n$ diode becomes a $p$--$n$--$n^+$ diode. Finally, the blank regions on the right upper part represent an inaccessible parameter space region for SiC, as the field exceeds the dielectric breakdown, $E_z>-400$~MV/m.

\subsection{Theory of the fluctuating electric field}

In addition to the static dc electric field $\textbf{E}\left(z_{\rm{def}},V\right)=(0,0,E_z(z_{\rm{def}},V))$ [Eq.~(\ref{elecf})] that leads to the shift of the defect frequency levels and optical transition energies, we also need to take into account  the temporally fluctuating electric field $\delta\textbf{E}\left(t\right)$ that makes the frequency levels fluctuate around the average frequency values dictated by $\textbf{E}\left(z_{\rm{def}},V\right)$. These fluctuations of the frequency levels produce a finite linewidth of the PL emission, an effect known as spectral diffusion. For the full ionized case of donors and acceptors, quasi-uniform electron and hole gases form due to the minimization of the Coulomb energy. This quasi-uniformity arises from various factors, e.g., particles' positions uncertainty (Heisenberg uncertainty principle), thermal fluctuation of electrons' and holes' positions, collision between different electrons (holes), etc. In addition, due to  thermal fluctuations, electrons (holes) can change from being free in the uniform gas, to becoming trapped on the donors (or acceptors) atoms, which is illustrated on Fig.~\ref{fig4}(a). Although someone could argue that these thermal fluctuations are not large for the $T\approx 10$~K of Ref.~\cite{anderson2019electrical}, the laser illumination used to address the defect PL raises the electronic temperature, thus making the thermal fluctuation a potential contributor to the fluctuating charge dynamics. We consider these effects by expressing the effective coupling of a general ground (excited) state level to the total (time-dependent) electric field $\textbf{E}\left(z_{\rm{def}},V,t\right)={\textbf{E}\left(z_{\rm{def}},V\right)}+\delta\textbf{E}\left(t\right)$
\begin{equation}
\frac{{\cal H}_{G\left(E\right)}}{h}=f_{G\left(E\right)}+\textbf{d}_{G\left(E\right)}^{\rm{eff}}\cdot\left[\textbf{E}\left(z_{\rm{def}},V\right)+\delta\textbf{E}\left(t\right)\right].
\end{equation}

Although a complete description of the quasi-uniform electron and hole gases is in principle a  correlated many-body problem, we treat the electrons and holes as  particles that do not interact with each other, 
due to their average separation  $l\approx n^{-\frac{1}{3}}\approx 46$~nm for typical diode $n$ carrier densities of $10^{16}$~cm$^{-3}$. More specifically, in this work we develop a theory for the fluctuating electric field $\delta\textbf{E}\left(t\right)$ using the physical process described in Fig.~\ref{fig4}. We solve this problem analytically, deriving closed form expressions for the fluctuating electric field as a function of the diode densities, diode dimensions and spin center's position. Furthermore, we see that these results agree very well with the experimental PL data of Ref.~\onlinecite{anderson2019electrical} and the numerical results from a Monte Carlo simulation, where we have fixed the donors' positions and build an electric field histogram by randomly changing the electronic positions. 


\subsection{Analytical calculation for the fluctuating electric field}

\begin{figure}[t!]
\begin{center}
\includegraphics[clip=true,width=1.0\columnwidth]{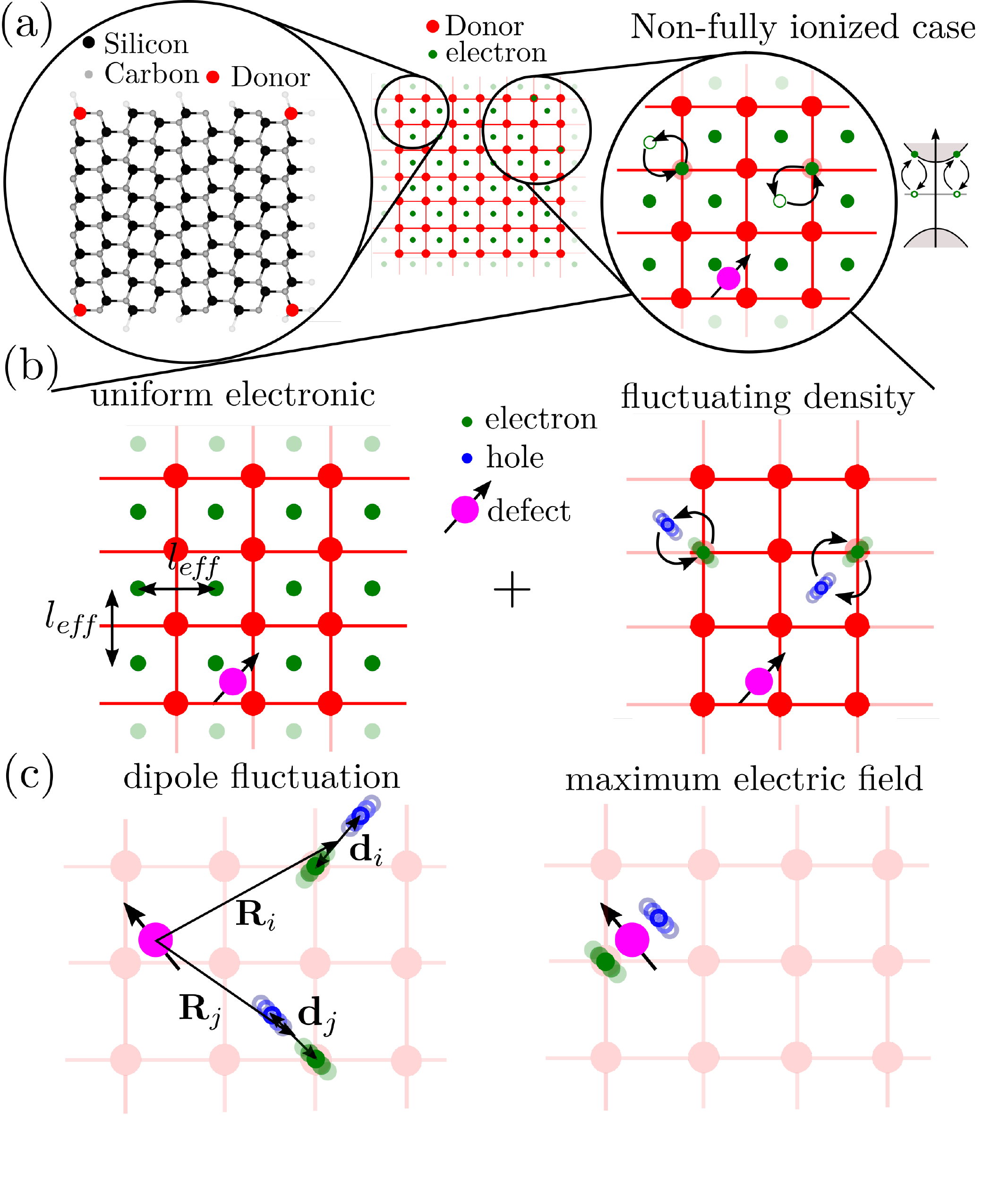}
\caption{(a) Schematic view of the 4H-SiC crystal structure with donors, together with the non-fully ionized situation with fluctuation of trapped charges. (b) The total electronic density is understood as being a sum of an uniform density, plus a fluctuating density of electron-hole dipole pairs. (c) We model the electron-hole pair fluctuating density though the dipole approximation, where the situation of maximum fluctuating electric field is shown.} 
\label{fig4}
\end{center}
\end{figure}

Fig.~\ref{fig4} describes the electronic structure of a semiconductor doped with donors to illustrate the origins of the fluctuating field. The schematically-indicated system is  4H-SiC, however the general approach is applicable to other semiconductor hosts. Carriers may not be fully ionized from the dopant atoms, and depending on the ionization fraction and other material parameters will produce a spatially fluctuating charge that we model here. As the positions of the charge fluctuations within Fig.~\ref{fig4} are random, we assume that all of the three components of the total electric field follow a Gaussian distribution, and will produce a linewidth $\Gamma$ of any specific optical transition. 
We consider optical emission associated with a transition from the ES to the GS (Fig.~\ref{fig1}) for a spin center at $z_{\rm{def}}$; the probability function $P\left(f\right)$ of emission of a photon with frequency $f$ is then
\begin{align}
P\left(f,z_{\rm{def}},V\right)  =\frac{1}{({2\pi})^{1/2}\Gamma}e^{-\frac{\left[f-\bar{f}(z_{\rm{def}},V)\right]^{2}}{2\Gamma^2}},
\label{emiss-spec}
\end{align}
with frequency emission peaked at
\begin{equation}
   \bar{f}(z_{\rm{def}},V)=f_{E}-f_{G}+\left(\textbf{d}_{E}^{\rm{eff}}-\textbf{d}_{G}^{\rm{eff}}\right)\cdot{\textbf{E}(z_{\rm{def}},V)},
\end{equation}
and our goal here is to calculate $\Gamma$.

We simplify the calculation of the linewidth by describing the fluctuation of charge density indicated within Fig.~\ref{fig4}(a) as being a sum of an uniform electron (and hole) density, plus a fluctuating dipole density --- as shown in Fig.~\ref{fig4}(b). Therefore, $\Gamma$ emerges from the standard deviation of a fluctuating dipole density. We calculate the $\delta\textbf{E}$  due to one instance $i$ 
of a dipole corresponding to the displacement of charges $e$ and $-e$ located at  $\textbf{r}_{i}\approx\left(x_{i},y_{i},z_{i}\right)$ 
and separated by the dipole distances $\textbf{d}_{i}$ [See Fig.~\ref{fig4}(c)].
Hence, the electric field at $\textbf{r}=\textbf{r}_{\rm{def}}$ produced by the {$i$}'th dipole is written as
\begin{align}
\textbf{E}_{d}^{i} \left( \textbf{R}_i \right) & =\frac{e}{4\pi\epsilon R_{i}^{5}}\left[3\left(\textbf{d}_{i}\cdot\textbf{R}_{i}\right)\textbf{R}_{i}-\textbf{d}_{i}R_{i}^{2}\right],
\label{dipole}
\end{align}
with $\textbf{R}_{i}=\textbf{r}_{i}-\textbf{r}_{\rm{def}}$ and $R_i=|\textbf{R}_{i}|$. The total fluctuating electric field then is
\begin{align}
|\delta\textbf{E}| =\sqrt{{\left\langle \textbf{E}_{d}^{2}\right\rangle_t} - {\left\langle \textbf{E}_{d}\right\rangle_{t}^2 }}, \label{contri-formal}
\end{align}
where $\textbf{E}_{d} = \sum_{i =1}^{N_{dip}} \textbf{E}_{d}^i\left( \textbf{R}_i \right)$ is the total electric field, and {$\left\langle \cdots \right\rangle_t $} represents the average in time over the different configurations (\textit{realizations}). Here, we assume $\left\langle \textbf{E}_{d} \right\rangle_t =0$ due to the large number of dipoles $N_{dip}\gg 1$, the random character of the considered fluctuations, i.e., $\left\langle \textbf{d}_{i}\right\rangle_t =0$ and $\left\langle {\textbf{R}}_{i}\right\rangle_t =0$, and that the random variables we introduce are  uncorrelated. We evaluate Eq.~(\ref{contri-formal}) assuming that the charge displacements $\textbf{d}_{i}$ $(d_i=|\textbf{d}_i|)$
are equally and randomly distributed along $x$, $y$ and $z$. Moreover, due to the random character of our variables and $N_{dip}\gg 1$, we choose to rewrite Eq.~(\ref{contri-formal}) using a continuous probability distribution for the dipolar position
\begin{equation}
    \delta\textbf{E}^2 = \int_{\mathcal{\mathcal{V}}} d^3r \rho_{\mathcal{V}}(\textbf{r}) \textbf{E}_{d}^2 (\textbf{r}-\textbf{r}_{\rm{def}}), \label{contri-tot}
\end{equation}
where $\mathcal{V}$ is the non-depleted volume region within the $p$--$n$--$n^+$ diode, and $\rho_{\mathcal{V}}(\textbf{r})$ the density of dipoles.

\begin{figure}[t!]
\begin{center}
\includegraphics[clip=true,width=1.0\columnwidth]{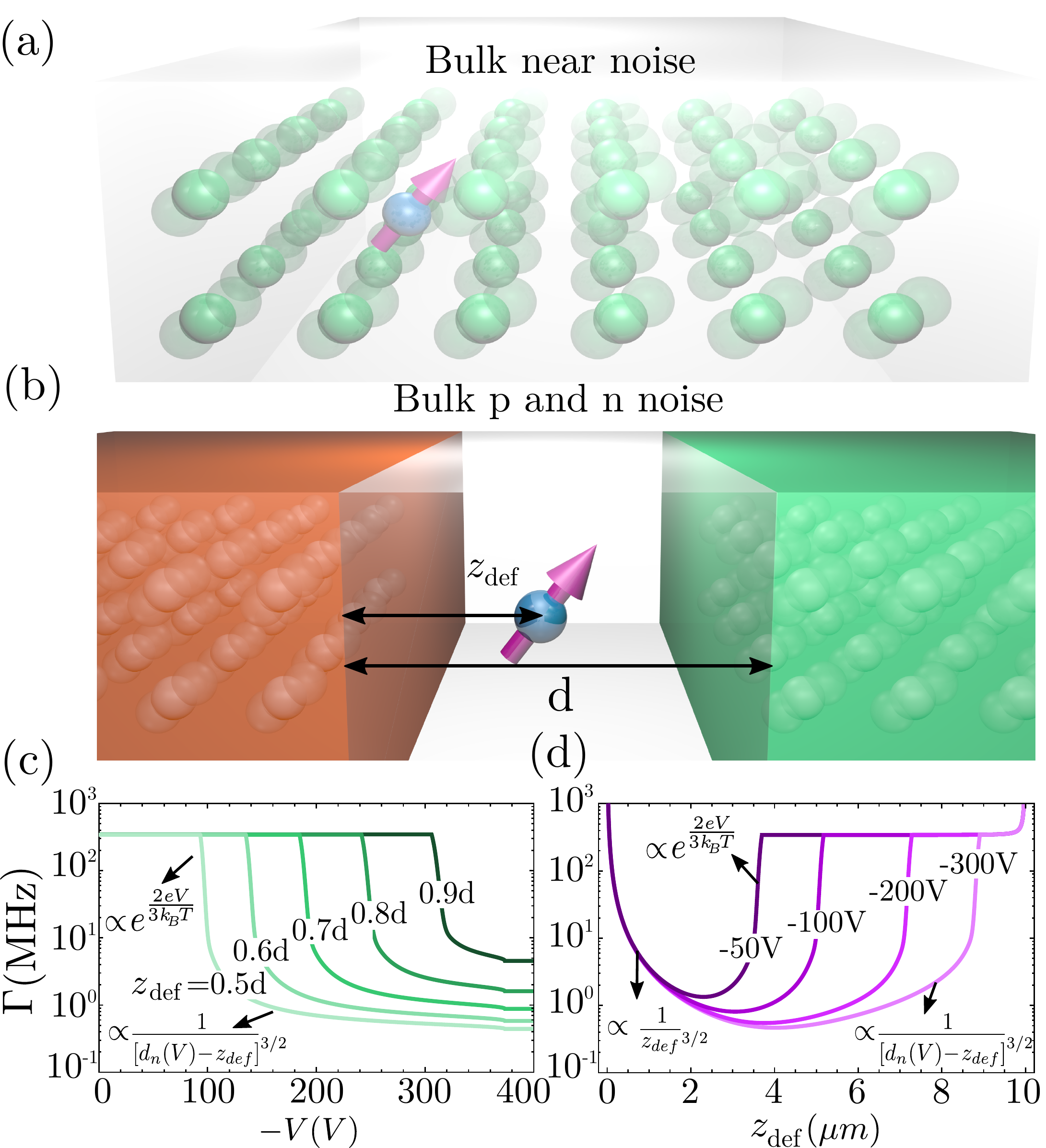}
\caption{Schematic view of the electric noise arising from the (a) \textit{n} region (bulk near noise contribution), (b) both \textit{p} and {\textit{n}}$^+$ regions (bulk \textit{p} and \textit{n} noise contribution). (c) Broadening $\Gamma$ as a function of reverse voltage for different $z_{\rm{def}}$ due to the both bulk near noise and bulk \textit{p} and \textit{n} contributions. (d) Same as (c) as a function of $z_{\rm{def}}$ for different voltage. } 
\label{fig5}
\end{center}
\end{figure}

Assuming there is no preferential direction for the total fluctuating electric field $|\delta\textbf{E}|$, we assume 
equal fluctuation of the electric field along the $x$, $y$ and $z$ axis, with nominal value along any one axis of $|\delta\textbf{E}|/\sqrt{3}$. For a linewidth  produced mainly due to the fluctuations of the $z$ component of the electric field, we then obtain
\begin{align}
\Gamma_{} = \frac{\left|\delta\textbf{E}\right|}{\sqrt{3}} \left( \textbf{d}_{E}^{eff}-\textbf{d}_{G}^{eff}\right)\cdot \hat{z}. \label{broadening}
 \end{align}
To obtain the realistic fluctuating electric field $\left|\delta\textbf{E}\right|$ for a device, we assume two different contributions to the electric noise. The first one, which we refer to as \textit{bulk near noise}, arises from the fluctuation of the electrons surrounding the defect in the \textit{n} region [Fig.~\ref{fig5}(a)]. The second types,  \textit{bulk \textit{p} and \textit{n} noise}, originate from the fluctuation of  electrons and holes within the \textit{p} and $n^+$ regions [Fig.~\ref{fig5}(b)]. As the spin centers are located far from any surfaces of the diodes, we do not consider sources of noise originating from the surfaces; this will be the topic of future work. In the following subsections we calculate analytically $\delta\textbf{E}$ [Eq.~(\ref{contri-tot})] arising from the different contributions illustrated in Fig.~\ref{fig5}(a) and (b).

\subsubsection{Fluctuating electric field: bulk near noise contribution}



In this subsection we estimate the fluctuating electric field $\delta\textbf{E}_{\mathcal{V}}^\textrm{I}$
at $\textbf{r}_{\rm{def}}=\left({x_{\rm{def}}},y_{\rm{def}},z_{\rm{def}}\right)$ produced by the fluctuation of trapped charges within
the non-depleted volume of the $n$ diode region [Fig.~\ref{fig5}(a)].  We evaluate Eq.~(\ref{contri-tot}) assuming charge displacements $\textbf{d}_{i}$ $(d_i=|\textbf{d}_i|)$
equally distributed along $x$, $y$ and $z$, with density $\rho_{\mathcal{V},\rm{I}}(\textbf{r})=1/{3\Omega_{\mathcal{V}}^n}$. For a non-depleted \textit{n} region we have 
\begin{align}
(\delta\textbf{E}_{\mathcal{V}}^{\rm{I}})^2 & =\left(\frac{e}{4\pi\epsilon}\right)^{2} \frac{6d_{i}^2}{3\Omega_{\mathcal{V}}^n}\int_{0}^{d}dz\int_{-\frac{L_{x}}{2}}^{\frac{L_{x}}{2}}dx\int_{-\frac{L_{y}}{2}}^{\frac{L_{y}}{2}}dy \frac{1}{R_{i}^6}. \label{cI-integral}
\end{align}
For diode with dimensions $d,L_x,L_y \gg N^{-1/3}$ and the spin center far from diode surfaces, we can extend the integral limits to infinity, yielding in spherical coordinates
\begin{align}
(\delta\textbf{E}_{\mathcal{V}}^{\rm{I}})^2 & =8\pi\left(\frac{e}{4\pi\varepsilon}\right)^{2}\frac{d_{i}^{2}}{\Omega_{\mathcal{V}}^n}\int_{r_{c}}^{\infty}dr\frac{1}{r^{4}} \label{cI-integral-sphe},
\end{align}
where the cutoff radius, $r_c$, was introduced to avoid the integral divergence at $r\rightarrow 0$. Using $d_{i}\approx l_{\rm{eff}}=\left(\Omega_{V}^{n} \right)^{\frac{1}{3}}$, $\Omega_{\mathcal{V}}^n=n_{}^{-1}$ and $r_c \approx n_{}^{-1/3}$, we obtain
\begin{equation}
|\delta \textbf{E}_{\mathcal{V}}^{\rm{I}} | \approx \frac{e}{\sqrt{2}\pi\varepsilon} n_{}^{2/3}(z_{\rm{def}},V). \label{Ebulknear}
\end{equation}
where $n(z_{\rm{def}},V)$ is the effective electronic carrier density within region I. Furthermore, there is an upper bound for the electric field, $E_{max}$, which correspond to the field in the middle of two opposite dipole charges $+e$ and $-e$,
separated by $\approx l_{\rm{eff}}$ [Fig.~\ref{fig4}(c)], 
\begin{align}
E_{max} & =\frac{e}{4\pi\epsilon\left(l_{\rm{eff}}/2\right)^{2}}+\frac{e}{4\pi\epsilon\left(l_{\rm{eff}}/2\right)^{2}}, \nonumber \\
 & =\frac{2e}{\pi\epsilon}n_{}^{2/3}(z_{\rm{def}},V). \label{Emax}
\end{align}
On the other hand, for a \textit{n} region partially depleted, we have 
\begin{align}
(\delta\textbf{E}_{\mathcal{V}}^{\rm{I}})^2 & =\left(\frac{e}{4\pi\epsilon}\right)^{2} \frac{6d_{i}^2}{3\Omega_{\mathcal{V}}^n}\int_{\tilde{d}_n\left(V\right)>z_{\rm{def}}}^{d}dz\int_{-\frac{L_{x}}{2}}^{\frac{L_{x}}{2}}dx\int_{-\frac{L_{y}}{2}}^{\frac{L_{y}}{2}}dy \frac{1}{R_{i}^6}. \label{cI-integral-a}
\end{align}
which for $L_x,L_y \gg N^{-1/3}$ yields 
\begin{align}
(\delta\textbf{E}_{\mathcal{V}}^{\rm{I}})^2 &=\left(\frac{e}{4\pi\epsilon}\right)^{2}\frac{d_{i}^{2}}{3\Omega_{\mathcal{V}}^{n}}\pi \times \nonumber \\
& \left\{\frac{1}{\left[\tilde{d}_n\left(V\right)-z_{\rm{def}}\right]^{3}}-\frac{1}{\left[\tilde{d}_n\left(V\right)-z_{\rm{def}}+d\right]^{3}} \right\}. \label{contriI}
\end{align}
Using $d_{i}\approx l_{\rm{eff}}=\left(\Omega_{V}^{n} \right)^{\frac{1}{3}}$, $\Omega_{\mathcal{V}}^n=N^{-1}$ and $d\gg \tilde{d}_{n}(V)-z_{\rm{def}}$, we obtain
\begin{align}
\delta\textbf{E}_{\mathcal{V}}^{\rm{I}} \approx \frac{e}{4\pi\epsilon}\frac{\sqrt{\pi}N^{1/6}}{\sqrt{3}} \frac{1}{\left[\tilde{d}_n\left(V\right)-z_{\rm{def}}\right]^{3/2}}, \label{contriI-approx}
\end{align}

\subsubsection{Fluctuating electric field: bulk $p$ and $n$ noise contribution}

We now calculate the fluctuating electric field at $\textbf{r}{_{\rm{def}}}$ due to the fluctuating of trapped charges within $n^+$ and $p$ regions [Fig.~\ref{fig5}(b)]. The procedure is very similar to the one approached in the previous subsection. The only difference regards the limit of the integration of Eq.~(\ref{cI-integral}). Here we have to integrate over the non-depleted $p$ and $n^+$ regions, yielding
\begin{align}
(\delta\textbf{E}_{\mathcal{V}}^{\rm{II}})^2 & =\frac{6d_{i,n^+}^2\left(\frac{e}{4\pi\epsilon}\right)^{2}}{3\Omega_{\mathcal{V}}^{n^+}}\int_{d_n(V)}^{d+d_R}dz\int_{-\frac{L_{x}}{2}}^{\frac{L_{x}}{2}}dx\int_{-\frac{L_{y}}{2}}^{\frac{L_{y}}{2}}dy \frac{1}{R_{i}^6} \nonumber \\
+& \frac{6d_{i,n^p}^2\left(\frac{e}{4\pi\epsilon}\right)^{2}}{3\Omega_{\mathcal{V}}^{n^p}}\int_{-d_{L}}^{-d_{p}\left(V\right)}dz\int_{-\frac{L_{x}}{2}}^{\frac{L_{x}}{2}}dx\int_{-\frac{L_{y}}{2}}^{\frac{L_{y}}{2}}dy \frac{1}{R_{i}^6}, \label{cII-integral}
\end{align}
where $d_{i,n^+}$ and $d_{i,n^p}$ are the dipole displacement within $n^+$ and $p$ regions, respectively, and $\Omega_{\mathcal{V}}^{n^+}$ and $\Omega_{\mathcal{V}}^{n^p}$ are the density of dipoles within $n^+$ and $p$ regions, respectively. Relying on the convergence of the integral we evaluate it using $L_x,L_y\rightarrow \infty$, leading to
\begin{align}
 (\delta\textbf{E}_{\mathcal{V}}^{\rm{II}})^2 &=  \left(\frac{e}{4\pi\epsilon}\right)^{2}\frac{d_{i,n^+}^{2}}{3\Omega_{\mathcal{V}}^{n^+}}\pi \times  \nonumber\\
 & \left\{\frac{1}{\left[d_n\left(V\right)-z_{\rm{def}}\right]^{3}}-\frac{1}{\left[d_n\left(V\right)-z_{\rm{def}}+d+d_R\right]^{3}} \right\} \nonumber\\
 & + \left(\frac{e}{4\pi\epsilon}\right)^{2}\frac{d_{i,n^p}^{2}}{3\Omega_{\mathcal{V}}^{n^p}}\pi \times \nonumber \\ 
 & \left\{\frac{1}{\left[d_{p}\left(V\right)+z_{\rm{def}}\right]^{3}}-\frac{1}{\left[d_{p}\left(V\right)+z_{\rm{def}}+d_{L}\right]^{3}}\right\}. \label{contriII}
\end{align}
Assuming the defect is closer to the $n^+$ side with $d_R\gg d$ and using $d_{i}\approx l_{\rm{eff}}=\left(\Omega_{\mathcal{V}}^{n^+} \right)^{\frac{1}{3}}$, with $\Omega_{\mathcal{V}}^{n^+}=N_{D}^{-1}$, we obtain
\begin{align}
 \delta\textbf{E}_{\mathcal{V}}^{\rm{II}} \approx \frac{e}{4\pi\epsilon}\frac{\sqrt{\pi}N_{D}^{1/6}}{\sqrt{3}} \frac{1}{\left[d_n\left(V\right)-z_{\rm{def}}\right]^{3/2}}. \label{contriII-approx}
\end{align}
This expression give us the important quantities to be controlled in order to produce diodes with reduced broadening of the optical transition energy due to fluctuating electric fields coming from distant regions.

\subsubsection{Voltage control of the optical emission linewidth}

Using the diode densities and dimensions of Fig.~\ref{fig2}, we plot in Fig.~\ref{fig5}(c) the broadening $\Gamma_{\rm{I+II}}$ [Eq.~(\ref{broadening})] due to the bulk near, \textit{p} and \textit{n} noise contributions [Eqs.~(\ref{contriI}) and (\ref{contriII})] as a function of the voltage. For the bulk near noise contribution, we assume that 3/4 of the trapped electrons within region \textit{n} are in deep traps,  and  therefore,  only  1/4  contribute  to  the  fluctuating electric field, i.e., $n_{}(z_{\rm{def}},V)=n_c(z_{\rm{def}},V)/4$. We plot $\Gamma_{\rm{I+II}}$ for different spin center position{s},  from $z_{\rm{def}}=0.5d$ to $0.9d$. For a fixed spin center position [Fig.~\ref{fig5}(c)], we  obtain a constant $\Gamma$ for voltages $|V|<|V_{th}|$. This correspond to the situation where the spin center is surrounded by undepleted carrier  electrons within \textit{n} region, and the optical emission linewidth is mainly due to the bulk near noise contribution [Fig.~\ref{fig5}(a)]. 

As we begin to increase the voltage magnitude $|V|\gtrsim |V_{th}|$, we deplete the electrons surrounding the spin center. When the depletion region reaches the spin center's position, $ n_{c}\approx e^{\frac{ eV}{k_{\rm{B}}T}}$ follows from Eq.~(\ref{nc}) and we obtain $\Gamma_{\rm{I}}\propto  e^{\frac{2 eV}{3k_{\rm{B}}T}}$, which is responsible for the exponential decay of the broadening in Fig.~\ref{fig5}(c). For voltages $|V_c|>|V|>|V_{th}|$, the \textit{n} region is still not fully depleted, although $ n_{c}(z_{\rm{def}},V)\approx0$. Hence we have the broadening due to Eq.~(\ref{contriI-approx}), $\Gamma_{\rm{I}} \propto \frac{1}{\left[\tilde{d}_n\left(V\right)-z_{\rm{def}}\right]^{3/2}}$, together with the broadening due to the bulk \textit{p} and \textit{n} noise contribution, $\Gamma_{\rm{II}} \propto \frac{1}{\left[d_n(V)-z_{\rm{def}}\right]^{3/2}}$, following from Eq.~(\ref{contriII-approx}).

For even larger voltages, $|V|>|V_{c}|$ the \textit{n} region becomes completely depleted, and therefore, the remaining broadening is  due to the bulk \textit{p} and \textit{n} noise contribution, $\Gamma_{\rm{II}} \propto \frac{1}{\left[d_n\left(V\right)-z_{\rm{def}}\right]^{3/2}}$. For our diode parameters,  $d_n(V)-d \ll d$ [Fig.~\ref{fig2}(d)] for $-1000<V<-100$~V, and an approximate independence with the voltage can be seen in Fig.~\ref{fig5}(c). In Fig.~\ref{fig5}(d) we plot $\Gamma_{\rm{I+II}}$ as a function of the spin center's position $z_{\rm{def}}$ for different reverse voltages, where  similar features can be seen.

\subsection{Monte Carlo simulation for the fluctuating electric field}

A numerical Monte Carlo simulation yields the fluctuating electric field at the spin center's position to compare with our analytic results. The results  validate the high degree of accuracy of our analytical approach. We numerically simulate the two types of noise contributions illustrated in Fig.~\ref{fig5}, using two different approaches. In the first one, a density $n$ ($p$) of the donors (acceptors) are assumed to have random and uncorrelated fixed positions, with electrons (holes) being randomly and uncorrelated placed among the entire considered region. In the second approach we account for the electrons' (holes') positions constrained within a sphere of radius $n ^{-1/3}$ around their correspondent donors' (acceptors') positions, thus capturing the dipole picture illustrated within Fig.~\ref{fig4}(b).

\subsubsection{Bulk near noise contribution}

\begin{figure}[t!]
\begin{center}
\includegraphics[clip=true,width=1.025\columnwidth]{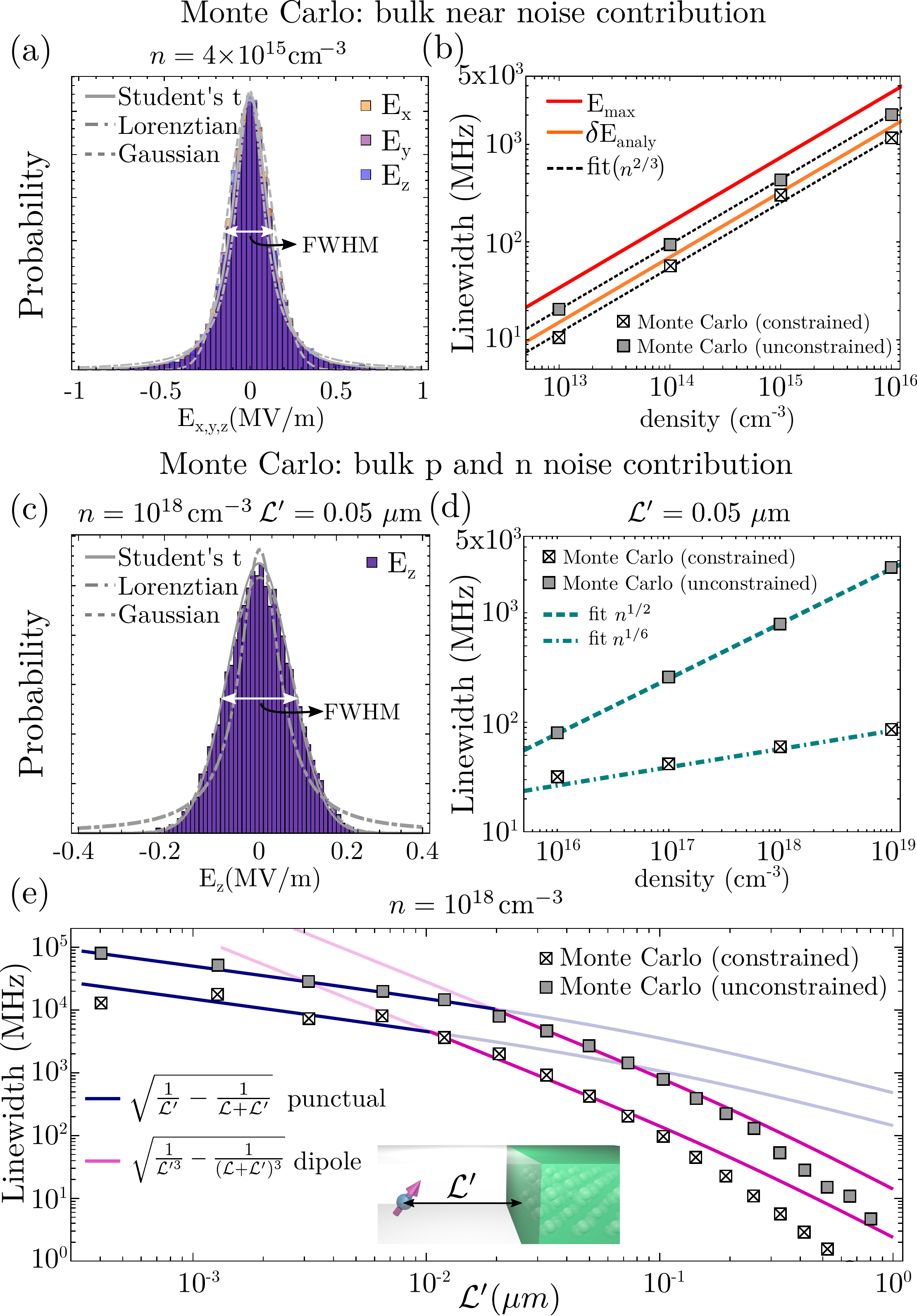}
\caption{(a) Electric field histogram due to the bulk near noise contribution with electronic density $n=4\times 10^{15}$~cm$^{-3}$. (b) Bulk near noise FWHM linewidth as a function of density $n$ for constrained and unconstrained electrons' positions. The dashed lines are fittings showing their $n^{2/3}$ density dependence, and the orange (red) is the plot of the analytical result Eq.~(\ref{Ebulknear}) (Eq.~(\ref{Emax})) (c) Electric field histogram due to the bulk $n$ noise contribution for electronic density $n=10^{18}$~cm$^{-3}$ and ${\cal L}'=0.05$~$\mu$m. (d) Bulk \textit{n} noise FWHM linewidth as a function of density $n$ for constrained and unconstrained electrons' positions for ${\cal L}'=0.05$~$\mu$m. (e) Bulk near \textit{n} noise FWHM linewidth as a function of ${\cal L}'$ for $n=10^{18}$~cm$^{-3}$  } 
\label{fig6}
\end{center}
\end{figure}

Here we obtain through Monte Carlo simulation the fluctuating electric field due to the bulk near noise contribution Fig.~\ref{fig5}(a). We proceed by assuming a spin center placed at the origin $\textbf{r}_{\rm{def}}=(0,0,0)$ of a box  with dimensions $\mathcal{L}\times\mathcal{L}\times\mathcal{L}$. $\mathcal{N}$ donors are randomly placed within our box (yielding a $n=\mathcal{N}/\mathcal{L}^3$ electronic density).  $\mathcal{N}$ electrons are then randomly placed accordingly to the two different approaches, and finally the total electric field at the spin center's position is calculated. A histogram for the three vector components of the total electric field at the spin center's position then is generated from a series of different electronic distributions in space (realizations). In Fig.~\ref{fig6}(a) we present the histogram for the density $n=4\times 10^{15}$~cm$^{-3}$, where we have used $\mathcal{N}=1000$ and $2\times 10^4$ different realizations. The best fit to the histogram is obtained for the Student's t-distribution rather than either the Gaussian distribution or the Lorenztian distribution. Roughly speaking, the Student's t-distribution differs from the Gaussian (Lorenztian) by its longer tail (broader peak region). This difference is clearly seen in Fig.~\ref{fig6}(a) where we have also fitted the data using the three different distributions. The underlying statistical reason for this requires more detailed study, but at this point we suggest this comes from  a) small number of nearby (influential) electrons  (small sampling size) and b) the electric field assumes large values when electrons are close to the defect, thus increasing the statistical weight of the tail of the distribution. Although this statistical distribution could be verified and realized experimentally, we note that in our Monte Carlo simulations the positions of electrons for different realizations have no correlation. Therefore, the realization of the Student's t-distribution may be done if the fluctuations in the electronic positions have short coherence  times. 

 Figure~\ref{fig6}(b) displays the full width at half maximum (FWHM) of the electric field histogram; the bulk noise contribution depends exclusively on the density $n$. The linewidth $\Gamma$ (FWHW) resulting from the random electric field distribution is plotted as a function of the density $n$ assuming constrained and unconstrained electronic positions relative to  their corresponding donors. For both cases the fit to the numerical data shows a{n}  $n^{2/3}$ dependence on the linewidth, which is explained by the analytical derivation presented in the previous section [Eq.~(\ref{Ebulknear})]. This shows that the density dependence of the bulk near noise is unlikely to be dependent on the details of the fluctuation character. In addition, we have also plotted the linewidth Eq.~(\ref{broadening}) arising from the Eqs.~(\ref{Ebulknear}) and (\ref{Emax}), where we  see good agreement between the analytical theory and Monte Carlo simulation.

\subsubsection{Bulk $n$ and $p$ noise contribution}
Here we study the fluctuating electric field due to the bulk \textit{p} and \textit{n} noise contribution Fig.~\ref{fig5}(b). We use the same procedure of the previous section, with the only difference of having now a spin center located outside our box, i.e., $\textbf{r}_{\rm{def}}=(0,0,\mathcal{L}'+\mathcal{L}/2)$. In Fig.~\ref{fig6}(c) we plot the histogram for the z component of the electric field, $E_z$, for $n=10^{18}$~cm$^{-3}$ (${\cal N}=1000$ and ${\cal L}=0.1$~$\mu$m) and ${\cal L}'=0.05$~$\mu$m. No longer is a long distribution tail visible, which is consistent with having the spin center far from the electrons, thus imposing an upper bound to the maximum electric field at $\textbf{r}_{\rm{def}}$. The Lorenztian remains a poor fit for our electric field histogram, however the  differences between the Student's t- and Gaussian distributions become less noticeable. However, a close look at the maximum of the histogram reveals that the Student's t-distribution  still produces a better fit. 

In Fig.~\ref{fig6}(d) we plot linewidth of the electric field distribution as a function of the density, assuming constrained and unconstrained electrons' positions, for $2\mathcal{L}'={\mathcal{L}}=0.1$~$\mu$m. Unlike our expectations from  bulk near noise, here the constrained and unconstrained situations yields a different linewidth dependence on the density $n$; the constrained case yields the dependence $n^{1/6}$, whereas the unconstrained case yields a larger linewidth with $n^{1/2}$ dependence. We emphasize that for  constrained electrons, which captures the dipole character of the process described in Fig.~\ref{fig4}(b), we obtain the same density dependence as the analytical formula Eq.~(\ref{contriII-approx}). On the other hand, the $n^{1/2}$ dependence for the unconstrained case can be easily derived if instead of using Eq.~(\ref{contri-tot}) for a continuous probability of dipoles, we use it for a continuous probability of point electron and hole densities $\rho_p(\textbf{r})$, with electric field $\textbf{E}_{p}(\textbf{R})=\pm\frac{e}{4\pi \epsilon}\frac{\textbf{R}}{|\textbf{R}|^3}$ i.e., 
\begin{equation}
    \delta\textbf{E}_{p}^2 = 2\int_{\mathcal{\mathcal{V}}} d^3r \rho_{p}(\textbf{r}) \textbf{E}_{p}^2 (\textbf{r}-\textbf{r}_{\rm{def}}). \label{contri-punc-int}
\end{equation}
Assuming again the $x$ and $y$ integration limits to be taken to infinity, we obtain for the bulk \textit{n} noise contribution 
\begin{equation}
|\delta\textbf{E}_{p}|=\frac{e}{\sqrt{2\pi} \epsilon}n^{1/2} \sqrt{\frac{1}{{\cal L}'}-\frac{1}{{\cal L}+{\cal L}'}},
\label{contri-punc}
\end{equation}
which agrees with the Monte Carlo results in Fig.~\ref{fig6}(d). Finally, the smaller linewidth produced by the constrained situation is traced to the smaller electric fields from the dipole field compared to that of point charges. 

In Fig.~\ref{fig6}(e) we plot the linewidth dependence on ${\cal L}'$ for both constrained and unconstrained electrons' positions with $n=10^{18}$~cm$^{-3}$. For the case of constrained electrons, the linewidth shows a $\sqrt{\frac{1}{{\cal L}'^3}-\frac{1}{({\cal L}+{\cal L}')^3}}$ dependence for ${\cal L}'> 0.1~\mu$m, which agrees with our analytical derivation for dipoles, Eq.~(\ref{contriII}). On the other hand, for ${\cal L}'< 0.1~\mu$m, the numerical data shows a dependence with $\sqrt{\frac{1}{{\cal L}'}-\frac{1}{{\cal L}+{\cal L}'}}$ instead, which is a scaling characteristic of the point charge distribution as shown by Eq.~(\ref{contri-punc}). Although this may appear to contradict the constrained character of the simulation, the dipole character only manifests at distances ${\cal L}'$ much larger then the dipole distance $n^{-1/3}\approx 0.01~\mu$m, and therefore, a dependence according to the expression for point charges  is expected for small values of ${\cal L}'$. Unconstrained electrons  show the same scaling behavior as the constrained ones, with the only difference a different overall constant factor. Since there is no constraint between electrons and donors, for small values of ${\cal L}'$ we do obtain the scaling $\sqrt{\frac{1}{{\cal L}'}-\frac{1}{{\cal L}+{\cal L}'}}$ of point charges, which agrees with our theory Eq.~(\ref{contri-punc}). However, for larger ${\cal L}'$ distances, the linewidth also scales with the dipole $\sqrt{\frac{1}{{\cal L}'^3}-\frac{1}{({\cal L}+{\cal L}')^3}}$ form [Eq.~(\ref{contriII})] even though  was no  constraint on electron position was imposed. Thus the dipole assumption used in the previous analytical section  provides a good picture for the effects of  charge noise on the linewidth. {Finally, for spin centers farther away than  the box dimension ${\cal L}$,  i.e., ${\cal L}'>0.1~\mu$m, the infinity limits taken on the $x$ and $y$ integration are not valid anymore, and a deviation from the analytical curve is seen.}

\subsection{Photoluminescence frequency and linewidth}

\begin{figure}[t!]
\begin{center}
\includegraphics[clip=true,width=1.0 \columnwidth]{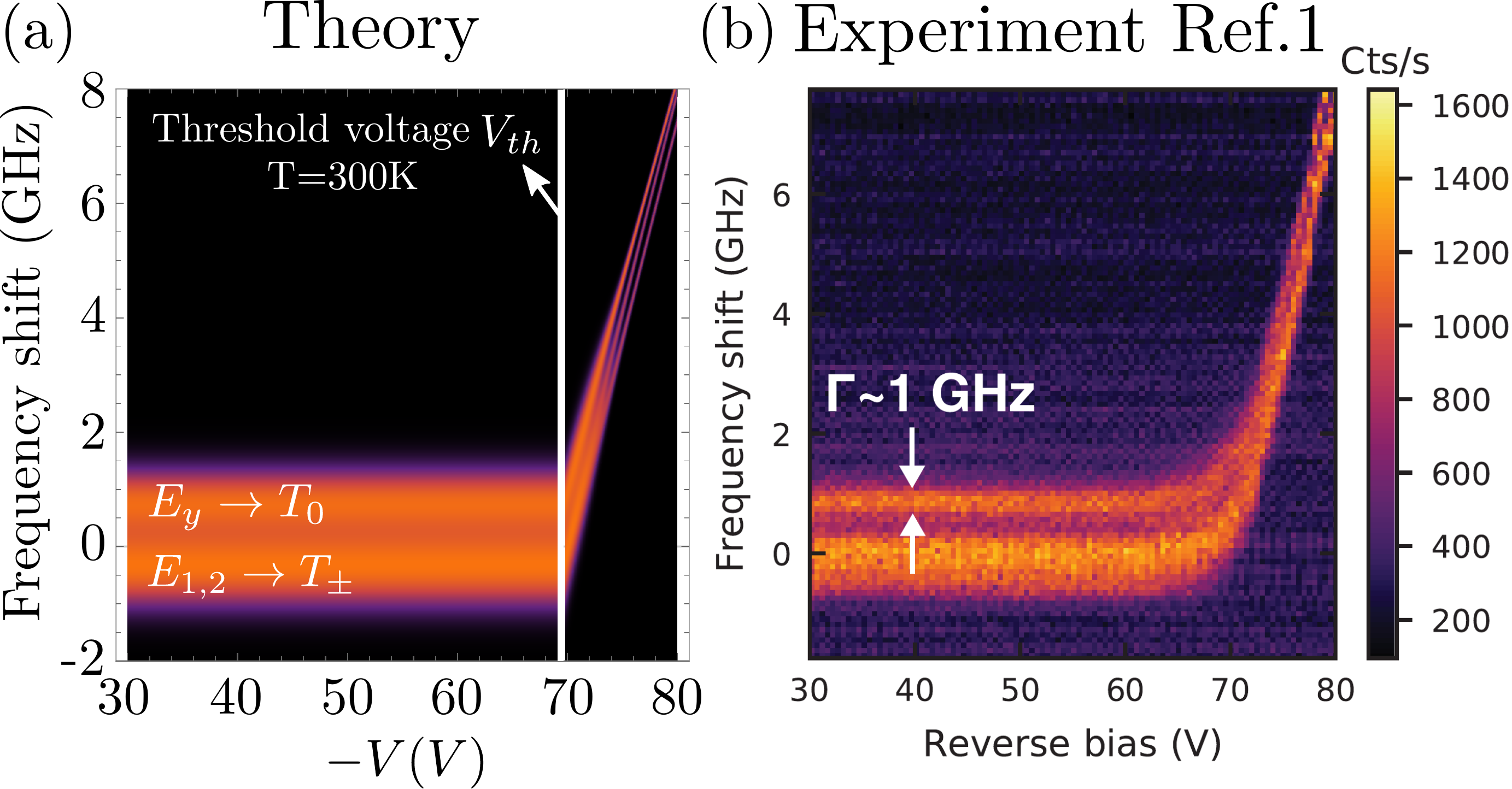}
\caption{Comparison between the theory (a) and experimental data from Ref.~\onlinecite{anderson2019electrical} (b) of a single (kk) divacancy photoluminescence with $\left|E_{y}\right\rangle  \rightarrow \left|T_{0}\right\rangle $  and $\left|E_{1,2}\right\rangle  \rightarrow \left|T_{\pm}\right\rangle$ addressed transitions. For $|V|<|V_{th}|$ the (kk) defect is within the non-depleted  part of the \textit{n} region, thus experiencing a null electric field and large electric noise and linewidth. For $|V|>|V_{th}|$  the charges begin to be depleted around the spin center, yielding a narrowing of the linewidth, and a shift in frequency in response to the now present electric field.} 
\label{fig7}
\end{center}
\end{figure}

The full dependence of the linewidth with respect to the temperature, electronic and hole densities, voltages and the position of the defect is obtained through Eq.~(\ref{broadening}), with the fluctuating electric fields calculated through Eqs.~(\ref{contriI}) and (\ref{contriII}).
Therefore, using the frequency shift expressions [Eqs.~(\ref{To-e})--(\ref{ao-ao})], together with the predicted linewidth [Eq.~(\ref{broadening})] we have a full theoretical characterization of the photoluminescence, including linewidth, from a divacancy  represented by Eq.~(\ref{emiss-spec}). In Fig.~\ref{fig7}(a), we plot the calcualted PL emission as a function of the  voltage for the (kk) defect of Fig.~\ref{fig3}(a). In our theoretical plot we consider both $\left|E_{y}\right\rangle \rightarrow \left|T_{0}\right\rangle $ and $\left|E_{1,2}\right\rangle  \rightarrow \left|T_{\pm}\right\rangle$ (kk) transitions. The corresponding experimental PL data of these transitions (Ref.~\onlinecite{anderson2019electrical}) is shown in Fig.~\ref{fig7}(b). Here, we have the (kk) defect located at $z=4.42~\mu$m (corresponding to $V_{th}\approx -70$~V and $N=4\times 10^{15}$~cm$^{-3}$), with $d_{E}^{\parallel}-d_{G}^{\parallel}=3.25$~GHz/(MV/m) and $E_{E_y}-E_G=-0.8$~GHz for the $\left|E_{y}\right\rangle \rightarrow \left|T_{0}\right\rangle $ transition and $d_{E}^{\parallel}-d_{G}^{\parallel}=3.55$~GHz/(MV/m) with $E_{E_1}-E_{G}=0.6$~GHz and $E_{E_2}-E_{G}=0.1$~GHz for the $\left|E_{1,2}\right\rangle  \rightarrow \left|T_{\pm}\right\rangle $  transitions. We emphasize that these dipoles values differ from the ones fitted in Fig.~\ref{fig3} due to the smaller fitting voltage range of Fig.~\ref{fig7}. As before we assume that only 1/4 of the carriers contribute to the fluctuating electric field due to deeper trapping effects, i.e., $n_{\rm{eff}}(z_{\rm{def}},V)=n_c(z_{\rm{def}},V)/4$. For voltages $|V|<|V_{th}|$ the spin center is outside the depletion region, thus experiencing zero electric field and hence a zero frequency shift, in addition to a large electric noise -- that leads to a large linewidth $\Gamma\approx 0.5$~GHz. For $|V|>|V_{th}|$ the carriers surrounding the spin center start to get depleted, allowing the spin center to experience a non-zero  electric field and a smaller fluctuating electric noise from the fewer carriers, leading to a shifted optical emission  frequency and a narrower linewidth. Due to the large diode dimensions $L_{x,y}\approx 100$~$\mu$m in Ref.~\onlinecite{anderson2019electrical},  noise  from the surfaces does not play any role and it is neglected in Fig.~\ref{fig7}. A very smooth linewidth transition around $V_{th}$ is found in the experimental data. A good agreement between theory and experiment is found by assuming an electronic temperature around the defect $\sim 300$K, produced by the laser illumination. The larger the temperature, the larger the tail of electronic density around the defect [$n_c(z_{\rm{def}},V\approx V_{th}) \propto e^{-\frac{eV}{k_B T}}$], which is the underlying reason for having the smooth linewidth transition as $\Gamma_I \propto n_c^{2/3}\approx e^{-\frac{2}{3}\frac{eV}{k_B T}}$ -- see Fig.~\ref{fig4}(g). 

\section{Spin decoherence  due to  electric noise}
\label{deco-sec}

\begin{figure}[t!]
\begin{center}
\includegraphics[clip=true,width=1.0\columnwidth]{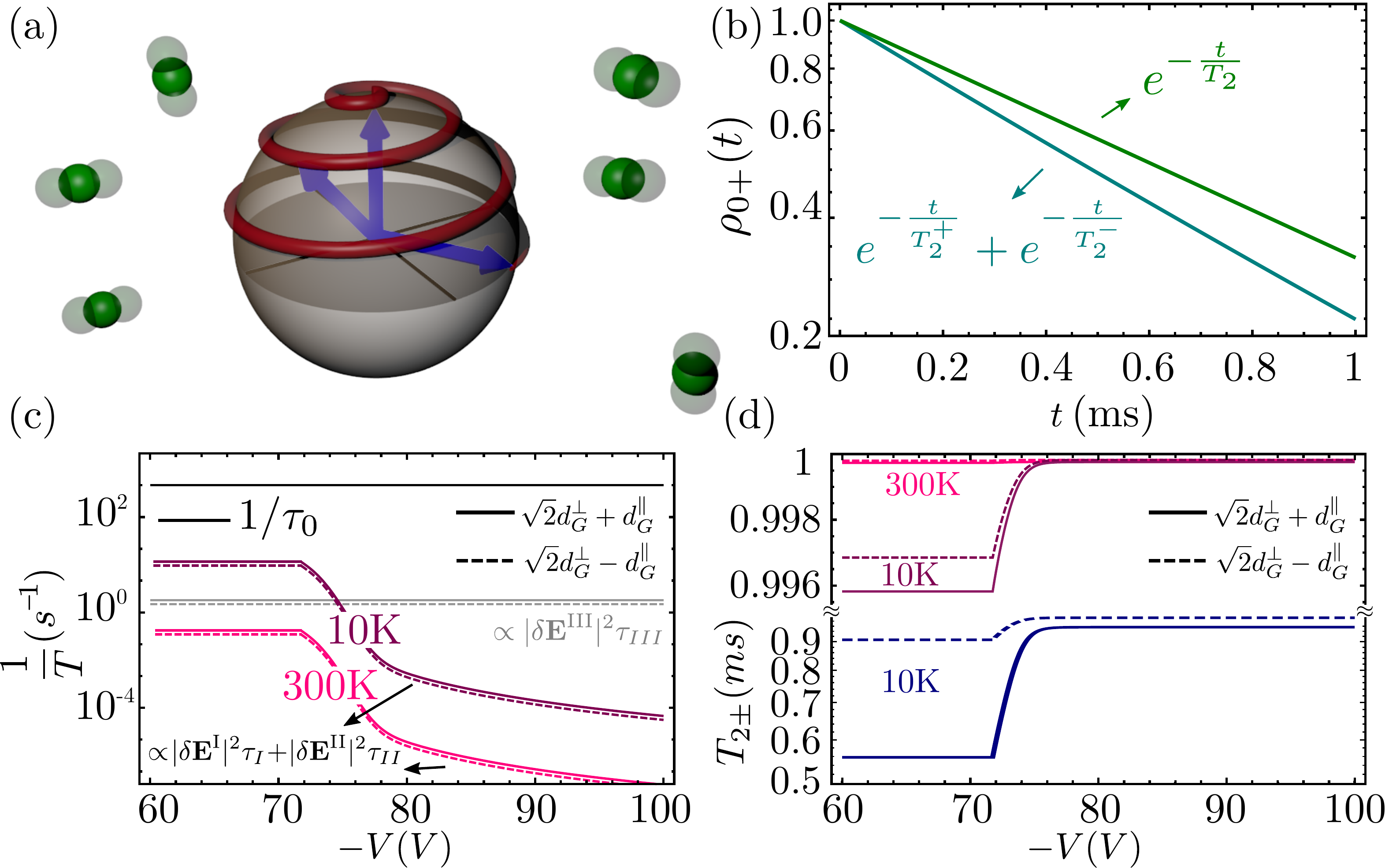}
\caption{(a) Schematic spin dephasing leading to decoherence due to the fluctuating electric charges. (b) Comparison between single and bi-exponential decoherence processes. (c) Decoherence contributions within Eq.~(\ref{T2pm}) as a function of diode voltage. (d) Enhancement of $T_{2\pm}$ decoherence times as a function of voltage due to the depletion of the electric noise. The dark blue curve is plotted exaggerating the $d_{\rm{G}}^{\parallel},d_{\rm{G}}^{\perp}$ ($\tau^{\rm{I}}$) dipole values by a factor of $\sim$10 (100). In (c) and (d) the solid (dashed) lines represent the contributions within $T_{2\pm}$ proportional to $\sqrt{2}d_{\rm{G}}^{\perp}+d_{\rm{G}}^{\parallel}$ ($\sqrt{2}d_{G}^{\perp}-d_{G}^{\parallel}$).} 
\label{fig8}
\end{center}
\end{figure}

Spin decoherence processes produce of a continuous loss of the memory of an initial state due to the influence and interaction with an environment. In this section we study the decoherence  of a single defect spin within a diode device. For our case, the fluctuating electric field produces --- in addition to the emission spectrum linewidth we have evaluated already  --- the spin decoherence  [See Fig.~\ref{fig8}(a)]. The random fluctuating electric fields produce a set of random phases in the wave function of the spin state. After averaging these random phases, we obtain a corresponding exponential temporal decay  of the state amplitude. We find that for defects with $C_{3v}$ point group symmetry, the precise spin decoherence process can only be addressed correctly through a $3\times 3$ spin-1 formalism, which includes the whole GS manifold. This in turn leads to a {\it bi-exponential} decoherence process that cannot be obtained through the usual $2\times 2$ spin 1/2 formalism~\cite{magneticnoise1,magneticnoise2,electricnoise1,magneticnoise4,electric-magnetic2,electric-magnetic4}. 
{We note that the complexity of the spin--1 manifold has been discussed phenomenologically before (\textit{e.g.} Ref.~\onlinecite{magneticnoise6}), however here we describe how the structure and temporal behavior of this decoherence emerges from microscopic models of electric noise.
}

\subsection{Spin-1 formalism for the decoherence of defects with $C_{3v}$ point group symmetry}

Here we apply the already-evaluated time-dependent fluctuating field due to the fluctuations of charges illustrated in Figs.~\ref{fig4} and \ref{fig5}, yielding $\textbf{E}(z,t)=\textbf{E}(z)+\delta\textbf{E}(t)$, with average in time $\left\langle \delta\textbf{E}\left(t\right)\right\rangle_t =0$ and deviation $|\delta \textbf{E}|= \sqrt {\left\langle \delta \textbf{E}(t)^2 \right\rangle _t}$ [Eq.~(\ref{contri-formal})].
Therefore, the GS Hamiltonian of a defect with $C_{3v}$ symmetry can be written as a sum of time-dependent and time-independent terms, 
\begin{equation}
    {\cal H}_{\rm{GS}}^{3v}(t) = {\cal H}_{\rm{GS}}^{3v} + {\cal H}'(t),
    \label{htotal-t}
\end{equation}
where ${\cal H}_{\rm{GS}}^{3v}$ is given by Eq.~(\ref{hgs}) and
\begin{align} 
\frac{{\cal H}'(t)}{h} & =d_{G}^{\perp} \delta E_{x}(t)\left(S_{y}^{2}-S_{x}^{2}\right)+d_{G}^{\perp} \delta E_{y}(t)\left(S_{x}S_{y}+S_{y}S_{x}\right) \nonumber \\
 & +d_{G}^{\parallel}\delta E_{z}(t) \left(S_{z}^{2}-\frac{2}{3}\right)\label{ht}.
\end{align}
For NV centers and (hh) and (kk) di-vancancies oriented along the diode direction ($z$ axis), the only relevant component of the dc electric field is along $z$. For this case, the eigeinstates of $\mathcal{H}_{GS}$ are still given by $\left|T_{-}\right\rangle$, $\left|T_{0}\right\rangle$ and $\left|T_{+}\right\rangle$. In addition, since a magnetic field along the $z$ direction also enters in the diagonal elements of our Hamiltonian, the eigenstates do not mix, and therefore, the presence of a magnetic field along {$z$} can also be addressed.
We now consider a general initial coherent state $\left|\psi\right\rangle (t=0)=\left|\psi\right\rangle _{0}=a_-\left|T_{-}\right\rangle +a_0\left|T_{0}\right\rangle +a_+\left|T_{+}\right\rangle$, and let it evolve in time in the presence of the total Hamiltonian Eq.~(\ref{htotal-t}). After a time $t$, the initial state $\left|\psi\right\rangle _{0}$ is
\begin{align}
\left|\psi\right\rangle \left(t\right)
 & =e^{-\frac{i}{h} {\cal T}\int_0^t dt \left[{\cal H}_{GS}+{\cal H}'\left(t\right)\right]}\left|\psi\right\rangle _{0},\nonumber \\ 
 & = a_{-}(t)\left|T_-\right\rangle+a_{0}(t)\left|T_0\right\rangle+a_+(t)\left|T_+\right\rangle,
\end{align}
with time ordering operator ${\cal T}$, and corresponding density matrix  $\hat{\rho}\left(t\right)=\sum_{\mu,\nu=-,0,+}a_{\mu}^{*}\left(t\right)a_{\nu}\left(t\right)\left|T_{\nu}\right\rangle \left\langle T_{\mu}\right|$.
We now take the average in time $\left\langle...\right\rangle_t$ over $\hat{\rho}(t)$ considering the temporal fluctuations of the fluctuating electric charges illustrated in Fig.~\ref{fig5}, yielding
\begin{equation}
\left\langle\hat{\rho}\left(t\right)\right\rangle=\left(\begin{array}{ccc}
\rho_{--}(t) & \rho_{0-}(t) & \rho_{+-}(t)\\
\rho_{-0}(t) & \rho_{00}(t) & \rho_{+0}(t)\\
\rho_{-+}(t) & \rho_{0+}(t) & \rho_{++}(t)
\end{array}\right)\label{rho3x3},
\end{equation}
with $\rho_{\nu\mu}\left(t\right)=\left\langle a_{\mu}^{*}\left(t\right)a_{\nu}\left(t\right)\right\rangle_t$. In this work we present the analytical result for the particular case of special experimental relevance, corresponding to $a_-=0$. This yields the initial coherent state $\left|\psi\right\rangle _{0}= a_0\left|T_{0}\right\rangle +a_+\left|T_{+}\right\rangle$, with $a_0^2+a_+^2=1$ arising from the normalization. For this particular case, the $a_{\nu}(t)$ coefficients are given by 
{\small{}
\begin{align}
\frac{a_{0}\left(t\right)}{a_0} & =e^{-iE_{0}t}F\left(\frac{2d_{\rm{G}}^{\parallel}}{3},t\right),\label{a0p}\\
\frac{a_{-}\left(t\right)}{a_+} & =\frac{e^{-iE_{-}t}e^{i\frac{\pi}{4}}}{2}\left[F\left(\sqrt{2}d_{\rm{G}}^{\perp}-\frac{d_{\rm{G}}^{\parallel}}{3},t\right)-F\left(-\sqrt{2}d_{\rm{G}}^{\perp}-\frac{d_{\rm{G}}^{\parallel}}{3},t\right)\right],\label{amp}\\
\frac{a_{+}\left(t\right)}{a_+} & =\frac{e^{-iE_{+}t}}{2i}\left[F\left(\sqrt{2}d_{\rm{G}}^{\perp}-\frac{d_{\rm{G}}^{\parallel}}{3},t\right)+F\left(-\sqrt{2}d_{\rm{G}}^{\perp}-\frac{d_{\rm{G}}^{\parallel}}{3},t\right)\right],\label{app}
\end{align} }where we have defined $F\left(\gamma,t \right) \equiv e^{i\gamma {\cal T} \int_{0}^{t}dt'\delta E\left(t'\right)}$. To obtain these expressions we assume the fluctuating fields along $x$, $y$ and $z$, $\delta E_{x,y,z}(t)$,  all have the same statistical properties, e.g., mean,  deviation and correlation function, and hence $\delta E_{x,y,z}(t)\equiv \delta E(t)$. To calculate the matrix elements of Eq.~(\ref{rho3x3}), we  temporally average the  product $\left \langle F\left(\gamma,t \right)  F\left(\delta,t \right) \right  \rangle  = \left \langle e^{ i(\gamma+\delta){\cal T}\int_{0}^{t}dt'\delta E\left(t'\right)}  \right \rangle= \left \langle F\left(\gamma+\delta,t \right) \right  \rangle $. This average is obtained by rewriting the exponential as an infinite series, and computing the average of its individual components, i.e., 
\begin{align}
\left\langle e^{i\gamma {\cal T} \int_{0}^{t}dt'\delta E\left(t'\right)}\right\rangle  & =1+i\gamma\int_{0}^{t}dt'\overset{=0}{\overbrace{\left\langle \delta E\left(t'\right)\right\rangle }}\nonumber  \\
 + & \frac{\left(i\gamma\right)^{2}}{2!}\int_{0}^{t}\int_{0}^{t'}dt'dt''\left\langle \delta E\left(t''\right)\delta E\left(t'\right)\right\rangle +\cdots \label{expansion-exp}
\end{align}
To compute Eq.~(\ref{expansion-exp}) we require the correlation function of the total fluctuating electric field at different times, $S(\tau,\tau')=\left\langle \delta E\left(\tau\right)\delta E\left(\tau'\right)\right\rangle $, and assume temporal transitional symmetry $S(\tau'-\tau)=S(\tau,\tau')$. It is also sufficient to know the noise spectral density $S(\omega)$, which is related to $S(t)$ through the Fourier transform $S\left(\tau'-\tau \right)=\int d\omega S\left(\omega\right)e^{i\omega (\tau'-\tau)}$. Here we will assume the total spectral density is described by a sum of two different Lorenztian noise spectral densities, arising from the two different sources of fluctuating electric charges, such as those illustrated in Figs.~\ref{fig5}(a) and (b). Thus
\begin{equation}
    S\left(\omega\right)= \sum_{\eta}
    \frac{(\delta\textbf{E}^{\eta})^2}{3\pi}\frac{\tau_{\eta}}{1+\omega^{2}\tau_{\eta}^{2}},
\end{equation}
where the index $\eta$ stands for the different noise contributions, $\tau_{\eta}$ represents its corresponding correlation time, and $\delta\textbf{E}^{\eta}$ represents its corresponding fluctuating electric field, given by Eqs.~(\ref{contriI}) and  (\ref{contriII}). The correlation function thus read $S\left(t\right) =\frac{1}{3}\sum_{\eta}(\delta\textbf{E}^{\eta})^2 e^{-\frac{\left|t\right|}{\tau_{\eta}}}$ and yields for $t\gg\tau_{\eta}$~\cite{anderson1953exchange}
\begin{equation}
    \left\langle e^{i\gamma{\cal T}\int_{0}^{t}dt'\delta E\left(t'\right)}\right\rangle \approx e^{-t \frac{\gamma }{6} \sum_{\eta}\left( \delta\textbf{E}_{\eta}\right)^{2} \tau_{\eta}} \label{exp-app}.
\end{equation}
Finally, using Eq.~(\ref{exp-app}) we obtain the diagonal terms
 \begin{align}
\rho_{00}\left(t\right) & =a_{0}^{2},\label{rho00p}\\
\rho_{--}\left(t\right) & =\frac{a_{+}^{2}}{2}\left(1-e^{-\frac{t}{T_1}}\right),\label{rho--p}\\
\rho_{++}\left(t\right) & =\frac{a_{+}^{2}}{2}\left(1+e^{-\frac{t}{T_1}}\right),\label{rho++p}
\end{align}
with characteristic longitudinal decay time
\begin{equation}
    \frac{1}{T_1}=\frac{\left(2d_{G}^{\perp}\right)^2}{3} \sum_{\eta}  \left(\delta\textbf{E}^{\eta}\right)^2 \tau_{\eta} +\frac{1}{\tau_0},
\end{equation}
and the off diagonal terms
\begin{align}
\rho_{0-}\left(t\right) & =a_{0}^{*}a_{+}e^{i\frac{\pi}{4}}e^{i\omega_{0-}t}\left(e^{-\frac{t}{T_{2-}}}-e^{-\frac{t}{T_{2+}}}\right),\label{rho0-p}\\
\rho_{0+}\left(t\right) & =a_{0}^{*}a_{+}e^{i\frac{\pi}{4}}e^{i\omega_{0+}t}\left(e^{-\frac{t}{T_{2-}}}+e^{-\frac{t}{T_{2+}}}\right),\label{rho0+p}\\
\rho_{-+}\left(t\right) & =\frac{\left|a_{+}\right|^{2}}{4}\left(e^{-\frac{t}{T_{2a-}}}-e^{-\frac{t}{T_{2a+}}}\right)\label{rho-+p},
\end{align}
with $\omega_{\mu \nu}={E_\mu - E_\nu}$, and the following four transverse decay  times
\begin{equation}
\frac{1}{T_{2\pm}}=\frac{1}{6}\left(\sqrt{2}d_{\rm{G}}^{\perp}\pm d_{\rm{G}}^{\parallel}\right)^{2}\sum_{\eta}  \left(\delta\textbf{E}^{\eta}\right)^2 \tau_{\eta}+\frac{1}{\tau_{0}}, \label{T2pm}
  \end{equation}
and
\begin{equation}
   \frac{1}{T_{2a\pm}}=\frac{2}{3}\left(\sqrt{2}d_{\rm{G}}^{\perp}\pm\frac{2}{3}d_{\rm{G}}^{\parallel}\right)^{2}\sum_{\eta}  \left(\delta\textbf{E}^{\eta}\right)^2 \tau_{\eta}+\frac{1}{\tau_{0}}.
 \label{T2apm}
  \end{equation}

We emphasize that in Eqs.~(\ref{rho--p})--(\ref{rho-+p}), the intrinsic decoherence time $\tau_0$  was introduced in order to incorporate other spin decoherence mechanisms that were not taken into account in our approach, e.g., dephasing from the random hyperfine nuclear fields of the surrounding atoms~\cite{seo2016}. This is an important element to add since in the absence of $\tau_0$ our approach would permit an infinite decoherence time as $|\delta \textbf{E}^{\eta}|\rightarrow 0$. It is also important to emphasize that different from many references~\cite{magneticnoise1,magneticnoise2,electricnoise1,magneticnoise4,electric-magnetic2,electric-magnetic4}, here we do not find only one transverse decay characteristic time, but rather four different ones, given by Eqs.~(\ref{T2pm}) and (\ref{T2apm}). 

A further analysis of Eqs.~(\ref{rho0-p})--(\ref{rho-+p}) shows that $\rho_{-+}(t)$ has a  decay time $\sim 4 \times$ faster than $\rho_{0-}(t)$ and $\rho_{0+}(t)$ [$T_{2a\pm} \ll T_{2\pm}$], and therefore we assume $\rho_{-+}(t)\approx 0$. In addition to that, we see that $\rho_{0-}(t)$ [Eq.~(\ref{rho0-p})] is given by a difference of two exponentials with similar arguments, so we also assume $\rho_{0-}(t)\approx0$. With those two approximations, $\rho_{0+}(t)$ is the responsible term for the decoherence. Surprisingly this term contains {\it bi-exponential relaxation} -- rather a single exponential -- with characteristic times given by Eq.~(\ref{T2pm}). 

In Fig.~\ref{fig8}(b) we plot $\rho_{0+}(t)$ with $T_{2+} \neq T_{2-}$ and $T_{2+} = T_{2-} = T_2$. We observe that the bi-exponential deviates from the linear trend within the log plot, which could be easily seen in experiments. The bi-exponential also produces a faster decay as compared to the case with $T_{2+} = T_{2-} = T_2$. The unusual bi-exponential feature appears due to the presence of the $d_{\rm{G}}^\perp$ dipole term as $T_{2+}=T_{2-}$ follows for $d_{\rm{G}}^\perp=0$, thus recovering the usual single exponential decay arising from the $2\times 2$ formalism. 

The bi-relaxation feature emerges as a sum of two different decoherent processes. The first  happens due to the dephasing of the initial state through the diagonal Hamiltonian terms that are proportional to $d_{\rm{G}}^\parallel \delta E(t)$ [see Eq.~(\ref{ht})], while the second one happens due to the off diagonal $d_{\rm{G}}^\perp \delta E(t)$ terms, which couple the $T_+$ and $T_-$ subspaces. The presence of these off-diagonal terms allows for an additional dephasing process, in which the loss of the memory of the initial state happens between the coupled $T_+$ and $T_-$ subspaces. As a consequence of this coupling we observe the increase of the population of the $\left|T_{-}\right\rangle$ state, $\rho_{--}(t)$ [Eq.~(\ref{rho--p})], which increases in time solely due to the presence of $d_{\rm{G}}^\perp \neq 0$ within $T_1$. The $3\times3$ density matrix formalism for the spin-1 system is  necessary to obtain this bi-relaxation process. If we had excluded the $T_-$ state due to its absence in the initial state $\left|\psi\right\rangle _{0}$, we would not obtain the two longitudinal dephasing times nor the increase of the $\left|T_{-}\right\rangle$ population. Most importantly, for NV centers and 4H-SiC divacancies we have $d_{\rm{G}}^{\perp} \gg d_{\rm{G}}^{\parallel}$~\cite{falko2014}, thus   the decoherence time is dominated by the $d_{\rm{G}}^{\perp}$ term that is not present in the $2\times 2$ formalism. Moreover, the results of the $2\times 2$ spin 1/2 formalism can be easily recovered from our formalism when $d_{\rm{G}}^\perp=0$, which leads to $T_{2+}=T_{2-}$, $T_{2a+}=T_{2a-}$, and  $\rho_{0-}(t)=\rho_{-+}(t)=\rho_{--}(t)=0$. 

\subsection{Decoherence times  as a function of diode voltage}

To evaluate how the decoherence times obtained in the previous subsection respond  to the diode voltage via Eqs.~(\ref{T2pm}), the correlation times of the fluctuating electric fields,  $\tau_{\eta}$ are required. There are two different correlation times, associated with the bulk near noise and the bulk \textit{p} and n$^+$ noise constributions, shown respectively in Figs.~\ref{fig5} (a) and (b). To estimate those quantities, we assume $T=10$~K for the electron temperature within \textit{p} and {$n$}$^+$ regions (experimental lattice temperature in Ref.~\onlinecite{anderson2019electrical}), and either $T=10$~K or $T=300$~K for the electrons within the \textit{n} region, due to the laser illumination. The estimate for $\tau_{\eta}$ comes from the relation between the mobility $\mu$ and diffusion coefficient $\mathcal{D}$ for  electrons and holes. For 4H-SiC,  $\mu_{10K}\approx 15$~cm$^{2}$/(V.s) and $\mu_{300K} \approx 70$~cm$^{2}$/(V.s)~\cite{mobility-4HSiC}. Using now the relation $\mathcal{D}=\frac{\mu}{e} k_{B}T$ and assuming $\mathcal{D}\approx l_{\rm{eff}}^{2}\frac{1}{\tau_{\eta}}$, with distance between trapped centers given by $l_{\rm{eff}}\approx n_{\rm{eff}}^{-1/3}$, we can establish a relation between the effective charge densities $n_{\rm{eff}}$ and the correlation time, namely $\tau_{\eta}=\frac{e}{\mu k_B T}n_{\rm{eff}}^{-2/3}$. Finally, we obtain $\tau_{\rm{I}}^{10K}\approx 8$~ns and $\tau_{\rm{I}}^{300K}\approx 55$~ps for $n_{\rm{eff}}$ density of $1\times 10^{15}$~cm$^{-3}$, and $\tau_{\rm{II}}^{10K}\approx 16$~ps for \textit{p} and {$n$}$^+$ densities $\approx 10\times 10^{18}$~cm$^{-3}$.

In Fig.~\ref{fig8}(c) we plot the different contributions of $1/T_{2}^{\pm}$ [Eq.~(\ref{T2pm})] as a function of the diode voltage $V$ for the characteristic times estimated above. The solid (dashed) lines correspond to the term proportional to $\sqrt{2} d_{\rm{G}}^{\perp} + d_{\rm{G}}^{\parallel}$ ($\sqrt{2} d_{\rm{G}}^{\perp} - d_{\rm{G}}^{\parallel}$) within $1/T_{2}^{+(-)}$. Here we used the same diode parameters as in Fig.~\ref{fig7}(a), where the spin center is located at $z=4.42~\mu$m with corresponding $V_{th}\approx -70$~V. We have assumed a $\tau_0=1$~ms, consistent with recent reported values~\cite{seo2016}. Due to the dipole inequality $d_{\rm{G}}^{\perp} \gg d_{\rm{G}}^{\parallel}$ following from $d_{\rm{G}}^{\perp} \approx 30\times 10^{-2}$~Hz/(V/m) and $d_{\rm{G}}^{\parallel} \approx 3\times 10^{-2}$~Hz/(V/m)~\footnote{Due to the lack of experimental available data for (hh) and (kk) divacancies ground state dipoles, here we have used the dipole values corresponding to the (hk) and (kh) divacancies~\cite{falko2014}.}, we barely see a difference regarding the different contributions proportional to $\sqrt{2} d_{\rm{G}}^{\perp} \pm d_{\rm{G}}^{\parallel}$, thus yielding $T_{2+}\approx T_{2-}$ (and no evident bi-relaxation).  For $|V|<|V_{th}|$, the spin center is placed inside the depletion region thus experiencing a large electric bulk near noise that yields for a large  $(d_{\rm{G}}^{\perp}  \delta \textbf{E}^{\rm{I}})^2 \tau_I$ contribution. On the other hand, for $|V|>|V_{th}|$ the depletion region reaches the spin center, and we start seeing a suppression of the bulk near noise and a consequent decreasing of $( d_{\rm{G}}^{\perp} \delta \textbf{E}^{\rm{I}})^2 \tau_I$, similarly to the narrowing of the PL linewidth within Fig.~\ref{fig7}. However, due to the small characteristic times $\tau_\eta \lesssim 10$~ns, we see that the bulk noise contributions are 3--5 orders of magnitude smaller than ${\tau_0}^{-1}$, thus showing the electric noise is irrelevant for decoherence for these parameters. This can also be seen on Fig.~\ref{fig8}(d), where we plot $T_{2\pm}$ as a function of the voltage for 10~K and 300~K temperatures of the \textit{n} region, purple and pink color curves, respectively. The enhancement of the coherence time due to the depletion of the surrounding electric noise is very small.
Therefore, for the diode setup of Ref.~\onlinecite{anderson2019electrical}, we should not observe an enhancement of the coherence time after depleting the bulk near fluctuations. This result corroborates with the  measurements of Ref.~\onlinecite{anderson2019electrical}, which did not observe any enhancement of $T_2$ for $|V|>|V_{th}|$. 

However, we note that if the electric dipole constant is $\sim$10 times larger, or if the correlation time is $\sim$100 larger, we would observe a clear enhancement of the coherence time after depleting the region surrounding the spin center. This is shown by the dark blue curve in Fig.~\ref{fig8}(d), where we have assumed $d_{\rm{G}}^{\perp} \approx 300\times 10^{-2}$~Hz/(V/m) and $d_{\rm{G}}^{\parallel} \approx 200\times 10^{-2}$~Hz/(V/m). In addition, with the new used values for the dipole constants we also see a clear difference between $T_-$ and $T_+$, which in turn would lead to an evident bi-relaxation process.

\section{Conclusions}

We provide a thorough and complete analytic and numerical theoretical description of the optical and electronic properties of a spin center in the presence of a dc electric field and local charge depletion produced through a voltage applied across a $p$--$n$--$n^+$ diode. Our results are in good agreement with the experimental measurements of Ref.~\onlinecite{anderson2019electrical}, and guide a more detailed understanding of the structure and properties of the materials used in the diode.  The diode structure allows for precise  control of the spin center's optical emission (PL) frequencies. Analytical expressions for the spin center's  transition frequencies are obtained as a function of the applied voltage and the diode parameters, which allows us to extract not only the electric dipole constants but also the spin center's position within the diode. We propose practical 4H-SiC diodes parameters that would allow frequency shifts of the PL emission in the THz range without dielectric breakdown. Moreover, the creation of the depletion region around the spin center's position removes electric noise from charges near the spin center, thus narrowing the spin center's PL linewidth, which we calculate analytically and simulate numerically with similar results.  Finally, we introduced a spin-1 formalism for the decoherence process of a spin center's ground state that yields a bi-exponential spin decoherence from microscopic models of electric field noise, and explained why this has not yet been seen experimentally for spin centers in diodes, however for closely related systems it should be possible to both observe these features and improve the spin coherence time through local charge depletion.

\begin{acknowledgments}
We thank D. D. Awschalom, C. P. Anderson, A. Bourassa, P. E. Faria, G. D. Fuchs, S. R. McMillan, A. R. da Cruz,  T. de Campos, Kwangyul Hu, C. \c{S}ahin and B. S. C. Candido for useful discussions. This work is supported by the U.S. Department of Energy, Office of Basic Energy Sciences, under Award Number DE-SC0019250.
\end{acknowledgments}

\appendix

\section{Frequency shift for alternative densities}
\label{appendixc}

\begin{figure}[b!]
\begin{center}
\includegraphics[clip=true,width=1.0\columnwidth]{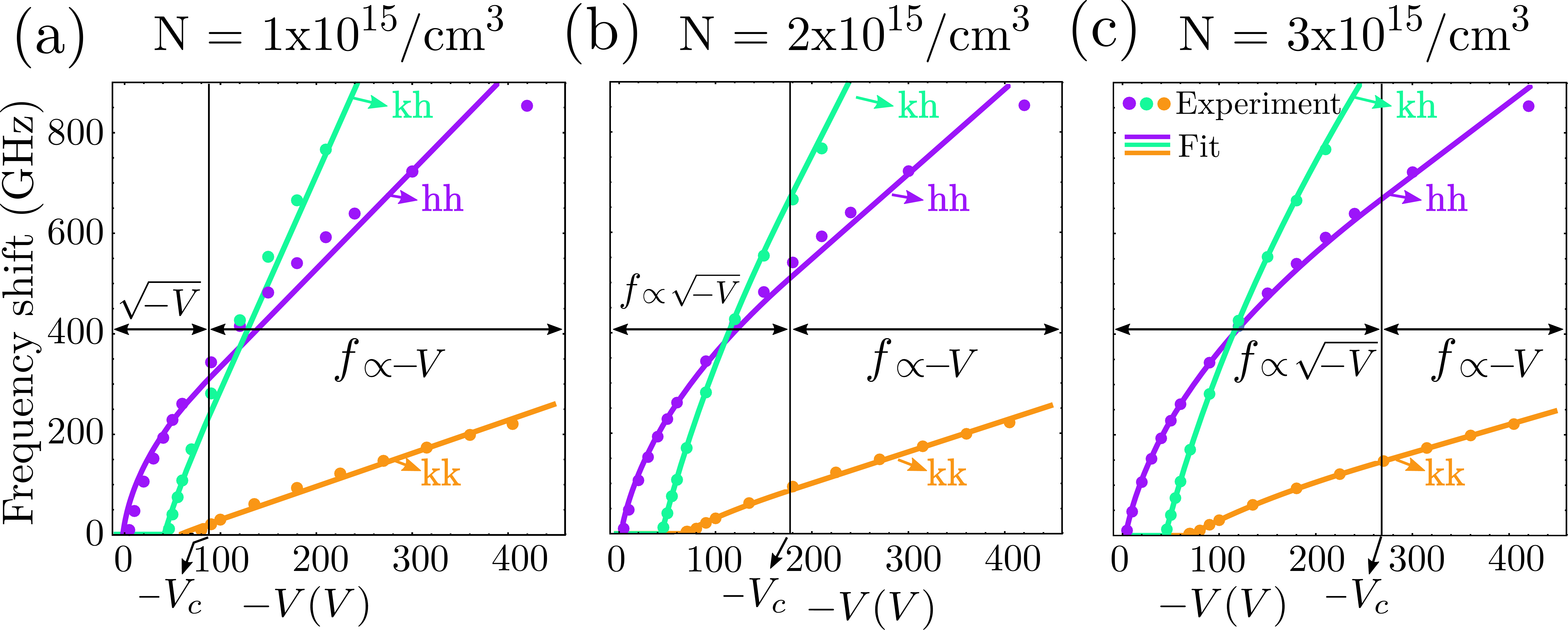}
\caption{ Fit for the frequency shift vs. voltage $V$ for the  $\left|E_{y}\right\rangle  \rightarrow \left|T_{0}\right\rangle$ and $ \left|E_{1,2}\right\rangle \rightarrow \left|T_{\pm}\right\rangle $ (hh) and (kk) transitions, and for the  $\left|A_{0}^{'}\right\rangle \rightarrow \left|\tilde{A}_{0} \right\rangle$ (kh) transition for densities (a) $N=1\times 10^{15}$~cm$^{-3}$, (b) $N=2\times 10^{15}$~cm$^{-3}$ and (c) $N=3\times 10^{15}$~cm$^{-3}$.} 
\label{figapp1}
\end{center}
\end{figure}

For completeness, in Fig.~\ref{figapp1} we also show the fit of the (kk), (hh) and (kh) divacancy frequency shifts within Fig.~\ref{fig3}, using densities (a) $N=1\times 10^{15}$~cm$^{-3}$, (b) $N=2\times 10^{15}$~cm$^{-3}$ and (c) $N=3\times 10^{15}$~cm$^{-3}$. It becomes evident that the larger the density $N$, the more accurately the experimental curve is fitted, thus suggesting  $N \gtrsim 3\times 10^{15}$~cm$^{-3}$. We propose that the fluctuating charges are the nominal charges in the $n$ region, whereas the additional charges are deeper traps due to the preparation properties of the material. These contribute to the depletion curves, however do not contribute to the optical linewidths due to their deep trap status.

\nocite{*}

\bibliography{apssamp}

\end{document}